\documentclass[useAMS,usenatbib]{mn2e}
\usepackage[]{txfonts}
\usepackage{threeparttable}
\def\simless{\mathbin{\lower 3pt\hbox
{$\rlap{\raise 5pt\hbox{$\char'074$}}\mathchar"7218$}}}   
\def\simmore{\mathbin{\lower 3pt\hbox
{$\rlap{\raise 5pt\hbox{$\char'076$}}\mathchar"7218$}}}   
\newcommand{\eqb}{\begin{eqnarray}}
\newcommand{\eqe}{\end{eqnarray}}
\newcommand{\cc}{C.C.}
\newcommand{\mel}{m_{\rm e}}
\newcommand{\mpr}{m_{\rm p}}
\newcommand{\thobs}{\theta_{\rm obs}}

\newcommand{\eb}{\epsilon_{\rm B}}

\newcommand{\sth}{\sigma_{\rm T}}
\newcommand{\Bcr}{B_{\rm cr}}
\newcommand{\rb}{r_{\rm b}}
\newcommand{\rem}{r_{\rm em}}

\newcommand{\vb}{V_{\rm b}}

\newcommand{\nph}{n_{\gamma}}
\newcommand{\nphhat}{\hat{n}_{\gamma}}
\newcommand{\nelhat}{\hat{n}_{\rm e}}
\newcommand{\gc}{\gamma_{\rm c}}
\newcommand{\linj}{\ell_{\rm e}^{\rm inj}}

\newcommand{\eobs}{\epsilon_{\rm obs}}

\newcommand{\Dop}{\mathcal{D}}
\newcommand{\mobs}{\mu_{\rm obs}}
\newcommand{\tesc}{t_{\rm e, esc}}
\newcommand{\tcr}{t_{\rm cr}}
\newcommand{\ub}{u_{\rm B}}
\newcommand{\us}{u_{\rm syn}}
\newcommand{\lsyn}{\ell_{\rm syn}}
\newcommand{\lssc}{\ell_{\rm ssc, i}}
\newcommand{\uex}{u_{\rm ex}}
\newcommand{\rex}{r_{\rm ex}}
\newcommand{\Lex}{L_{\rm ex}}
\newcommand{\eex}{\epsilon_{\rm ex, \star}}

\newcommand{\fex}{f_{\rm ex}}
\newcommand{\lb}{\ell_B}
\newcommand{\lex}{\ell_{\rm ex}}
\newcommand{\nel}{n_{\rm e}}
\newcommand{\nt}{N_{\rm T}}
\newcommand{\dt}{\delta t}

\newcommand{\epeak}{\epsilon_{\rm p, obs}}

\newcommand{\ks}{k_{\rm s}}
\newcommand{\kf}{k_{\rm f}}
\usepackage{graphicx}
\usepackage{epsfig} 
\usepackage{epstopdf}
\usepackage{color}
\usepackage{ulem}

\title[The Inverse Compton Catastrophe]
{Spectral signatures of compact sources in the inverse Compton catastrophe limit}

\author[Petropoulou, Piran, Mastichiadis]
{M. Petropoulou$^{1,2}$\thanks{E-mail: mpetropo@purdue.edu  (MP)},
T. Piran$^3$, A. Mastichiadis$^{4}$\\
$^{1}$Department of Physics and Astronomy, Purdue University, 525 Northwestern
Avenue, West Lafayette, IN 47907, USA\\
$^{2}$Einstein Postdoctoral Fellow\\
$^3$Racah Institute of Physics, The Hebrew University, Jerusalem 91904, Israel\\
$^4$Department of Physics, University of Athens, Panepistimiopolis, GR 15783 Zografos, Greece}

\begin{document}
\date{Received / Accepted}
\pagerange{\pageref{firstpage}--\pageref{lastpage}} \pubyear{2013}

\maketitle

\label{firstpage}

\begin{abstract}
The inverse Compton catastrophe is defined as a dramatic rise 
in the luminosity of inverse Compton scattered photons.
It is described by a non-linear loop of radiative processes that sets in for high values of the electron compactness
and is responsible  for the efficient transfer of energy from electrons to photons, predominantly through inverse Compton  scatterings.
We search for the conditions that drive a  magnetized non-thermal  source to the inverse Compton catastrophe regime and study its multi-wavelength (MW)  photon spectrum. We develop a generic analytical framework and use numerical calculations as a backup to the analytical predictions. We find that the escaping radiation from a source  in the Compton catastrophe regime bears some unique features. The MW photon spectrum is a broken power law with a break at $\sim \mel c^2$ due to the onset of the Klein-Nishina suppression. The spectral index below the break energy depends on the electron and magnetic compactnesses
logarithmically, while it is independent of the electron power-law index ($s$).  The maximum radiating power emerges typically in the $\gamma$-ray regime,  at energies $\sim \mel c^2$ ($\sim \gamma_{\max} \mel c^2$ ) for $s>2$ ($s\lesssim 2$),where $\gamma_{\max}$ is the maximum Lorentz factor of the injected electron distribution. We apply the principles  of the inverse Compton catastrophe 
to blazars and $\gamma$-ray bursts using the analytical framework we developed, and show how these can be used to 
impose robust constraints on the source parameters.
\end{abstract} 
  

\section{Introduction} 
\label{intro}
{ 
The {\sl inverse Compton catastrophe} is a non-linear loop of processes that
is closely related to the synchrotron self-Compton (SSC) emission when
electron cooling is taken into account  \citep[e.g.][]{longairHEA}.  
If a magnetized, non-thermal emitting source,
is sufficiently compact, then it is possible that the energy
density of the radiated photons dominates over the magnetic one, which
causes the electrons to lose energy mainly by inverse Compton scattering rather than
synchrotron radiation. In such a case, a runaway process is possible:
low energy (e.g. radio) photons produced by synchrotron radiation are scattered to higher energies (e.g. in X-rays)
by the same relativistic electron population. As the energy density of these high energy photons
is larger than that of synchrotron photons, the electrons suffer even greater energy losses
by up-scattering them to even higher energies, e.g. in $\gamma$-rays. In turn,
these have greater energy density than the X-ray photons, and so on. 
This runaway process leads, on the one hand, to an increase of the electron cooling rate and,
on the other hand, shifts the radiated photon power from low to
high frequencies. In particular, the radiated power 
is shifted to the $\gamma$-ray regime, up to energies of the order of the maximum electron
energy. Pushing this idea to its limits, the non-linear loop of processes can lead
to complete and very fast electron cooling and from this, the
term {\sl inverse Compton catastrophe} is coined.

An important concept in the study of the inverse Compton catastrophe 
are the multiple up-scattered photon generations. One can
label the photon generations produced in the source according   
to how many scatterings they have undergone after production by the synchrotron process.
The synchrotron photons are defined as the zeroth order photon generation to be produced
in the source, which, in turn, serve as the seed photons for the first inverse Compton scattered
photon generation; similarly,
photons belonging to the $i$-th generation are the seed photons for the $i+1$ generation.
The production of such higher order photon generations can take place
even if the source is optically thin to Thomson scattering, and in this case 
the dominant loss of photons belonging to a certain generation (with $i >0$) is 
the escape from the source within a crossing time (see also \citealt{bjornsson00, tsangkirk07}).

The inverse Compton catastrophe
has been mainly studied in the context of compact radio sources \citep[e.g.][]{kellermann69, readhead94, bjornsson00, tsangkirk07}. 
It has been initially invoked to explain the observed upper limit in the distribution of brightness temperatures ($T_{\rm br}$) 
from powerful extragalactic radio sources \citep{kellermann69}. 
Later it was suggested that the observed cutoff in the $T_{\rm br}$ distribution was not related to the
inverse Compton catastrophe, since the latter sets an upper limit of $T_{\rm br}^{\rm cc}\sim 10^{12}$~K, which lies above
the observed cutoff \citep{readhead94}.
Yet, $T_{\rm br}^{\rm cc}$ remains a theoretical upper limit, 
against which the observed brightness temperatures, e.g. derived from the intra-day variability in radio cores
of blazars, are still tested (e.g. \citealt{fuhrmann08} and references therein).

The efficient cooling of non-thermal electrons due to inverse Compton scatterings has been also studied
in a different context, i.e. that of unmagnetized\footnote{A first analytical treatment of pair cascade
in the synchrotron self-Compton model can be found in \cite{ZL85}; the problem of pair cascades in magnetized
sources has been studied also numerically, later on,  by \cite{coppi92}.} 
sources with high optical depths for photon-photon absorption and Compton
scattering.
A large volume of literature   \citep[e.g.][]{fabian86, LZ87, Svensson87, zdziarski88, zdziarskietal90,svensson94}
deals with the problem of pair (saturated or not) cascades that are (i) initiated by absorption of high-energy photons on a soft 
(fixed) 
photon field and (ii) mediated by efficient Compton cooling of the pairs.
The main goal of these studies was to determine
the properties of the escaping radiation under the modification of non-thermal pair cascades and explain, among others,
the universality of the observed spectral index in the X-ray spectra of active galactic nuclei 
\citep[e.g.][]{bonomettorees71, rothschild83, kazanas84, zdziarskietal90, svensson94}.

Here we consider the production of multiple photon
generations due to inverse Compton scattering in magnetized, optically thin to Thomson scattering, sources, and in the regime where pair cascades
are negligible. The physical setup is similar to previous studies of the inverse Compton catastrophe \citep[e.g.][]{tsangkirk07}, yet
our main goals differ. These are summarized in the following:
(i)  study in a systematic way the transition
from linear synchrotron to non-linear SSC cooling under the presence of higher order up-scattered
photon generations; (ii) develop a general analytical framework that is easy to be used and can be applied to different types
of sources; (iii) determine analytically the conditions, such as the electron compactness,
above which the higher order SSC photon generations dominate the energetic output of the source;
(iv)  derive an analytical expression for the spectral index in the limit
where the multi-wavelength photon spectrum is determined by higher order photon generations; and (v) use
numerical calculations to back up our analytical predictions.}

This paper is structured as follows. We present our assumptions and our analytical treatment in Sect.~\ref{AR}
and continue in Sect.~\ref{NR} with a presentation of our numerical results and a comparison against the analytical predictions. 
In Sect.~\ref{astro} we present 
indicative applications of the inverse Compton catastrophe in the context of high-energy emitting
astrophysical sources. We conclude in Sect.~\ref{discussion}  with a discussion of our results.

\section{Analytical results}
\label{AR}
{ 
We adopt a rather abstract framework for the description of the source, according to which,
the emission region of the source is assumed to be embedded in an external (fixed) radiation field, to be 
homogeneous and magnetized, and to contain a population of relativistic electrons, which is 
constantly being replenished. In this framework, we will derive (under certain simplifying assumptions)
analytical expressions for the conditions leading the source in the inverse Compton catastrophe limit. 
For this purpose, we will take into account not only synchrotron cooling of electrons, but also 
cooling on synchrotron and higher order inverse Compton-scattered photons. 
In order to keep the analysis as general as possible, all of our results will be expressed in terms of three basic quantities:
the electron, the magnetic and the external photon compactnesses, where the compactness is a dimensionless measure
of the respective energy density\footnote{The (dimensionless) compactness is traditionally 
defined as $\sim \sth L/ 4 \pi \mel c^3 R$ , where $L$ 
and $R$ are the luminosity and size of the system  \citep{herterich74}. 
} and extensively used in studies of the emission from compact sources \citep[e.g.][]{ZL85, Svensson87,LZ87, mastkirk95}.}
\subsection{Assumptions}
We consider a homogeneous spherical region of radius  $\rb$ 
and volume $\vb = 4 \pi \rb^3/3$ that immersed in a  magnetic field
of strength $B$, and moves relativistically with Lorentz factor $\Gamma$. As
the relativistic motion of the emitting region will not become central to our analysis until much later, 
the only reference frame of interest, at this point, is the rest frame of the emitting region, where all the following 
calculations will be performed.

For the purposes of the analytical treatment, we assume that 
mono-energetic { relativistic} electrons with Lorentz factor $\gamma_0 {\gtrsim 3}$ are being
injected in the emission region with a constant rate $Q_0$, 
while injection of a more realistic power-law distribution will be considered
in the numerical calculations. We further assume that electrons
may physically escape from the emission region
at the (energy independent) characteristic timescale $t_{\rm e, esc} = \rb /c$. This is also 
equal to the typical timescale for adiabatic losses, at least for a spherical source expanding close to the speed of light. 
Electrons lose energy through synchrotron radiation and inverse Compton scattering on the internally produced
synchrotron photons (SSC) and possibly, on external photons (external Compton or EC). 
In order to keep the analysis as general as possible, we will not 
specify  the origin of the external radiation until much later (Sects.~\ref{luminosities} and \ref{astro}).  
The only property of the external photon field that is currently important for our analysis
is its energy density $\uex$ as measured in the rest frame of the emitting
region.

We also approximate the emissivity of synchrotron and inverse Compton scattering by a $\delta$-function centered at 
(dimensionless) photon energies $x_{\rm syn}=b\gamma_0^2$ and $x_{\rm ssc,i}=(4/3)^{i}b\gamma_0^{2(i+1)}$, respectively. Here, $x_{\rm ssc,i}$
is the energy of a photon that has been Compton up-scattered $i$-times, 
$b=B/\Bcr$ with $B$ being the comoving magnetic field strength and $\Bcr=4.4\times10^{13}$~G.
Finally, in the analytical treatment we consider only optically thin to synchrotron self-absorption (ssa) cases (see \citealt{tsangkirk07}, for
the role of ssa in Compton catastrophe cases). 

\subsection{Steady-state solutions of the electron kinetic equation}
The evolution of the electron distribution $\nel$ is described by the kinetic equation
\eqb
\frac{\partial \nel(\gamma, t)}{\partial t} + \frac{\nel(\gamma, t)}{\tesc} = Q_e(\gamma, t) + \mathcal{L}_e(\gamma, t) ,
\label{kinetic}
\eqe
where $Q_e$ and  $\mathcal{L}_e$ are the injection and energy loss operators for electrons, respectively. 
These are defined as
\eqb
\label{source}
Q_{\rm e} & = & Q_0\delta(\gamma-\gamma_0)H(t), \\
\mathcal{L}_e & = & \frac{4}{3}\frac{\sth c}{m_e c^2}\frac{\partial}{\partial \gamma}\left[\gamma^2 \nel(\gamma, t) u_{\rm tot}\right],
\label{loss}
\eqe
where $H(x)$ is the Heaviside function and $u_{\rm tot}$ is the total energy density. This is given by
\eqb
u_{\rm tot} & = &  \ub + \uex + \us +\sum_{i=1}^{\nt-1} u_{\rm ssc,i}.
\label{utot}
\eqe
The last term takes into account the energy 
density of successive SSC photon generations, starting from the 
first one ($i=1$) that is the result
of the synchrotron photon up-scattering ($i=0$) and terminating at the $(\nt-1)$ photon generation. 
These photons are,
in turn, the
seeds for the last SSC generation to be produced in the Thomson regime.
SSC photons belonging to the $\nt$-th generation appear in the electron's rest frame
with energy $\gg \mel c^2$, and thus, their up-scatter will 
take place in the Klein-Nishina regime,
where the cross section for scattering is greatly suppressed. This also explains the absence of $u_{\rm ssc, \nt}$
from the electron cooling term. The highest order of up-scattered photons in the Thomson regime, $\nt$, is given by
\eqb
\nt = \left[\frac{\log \left({1}/{b\gamma_0} \right)}{\log\left({4\gamma_0^2}/{3} \right)}\right],
\label{NT}
\eqe
where the square brackets denote the closest (from below) integer value of the enclosed expression (see Appendix~\ref{app1} for the derivation). 

In principle, one should also include in the equation of $u_{\rm tot}$ a term related to the 
repeated inverse Compton scatterings of external photons, similar to the last term of the r.h.s. in eq.~(\ref{utot}).
It can be shown, however, that for
$\gamma_0\sim 100$ and typical values for the energy of external photons (e.g. 1~eV), even the first inverse Compton scattering 
of the EC photons takes place deep in the Klein-Nishina regime (see Appendix~\ref{uEC}). In this case, one can 
safely ignore the term related to the energy density of EC photons. However, this does not apply if the injected electrons
are less energetic, e.g. their Lorentz factor is $\sim 10$. Thus, for completeness reasons, we derive in Appendix~\ref{uEC}
similar relations and constraints to these presented in the following paragraphs (Sects.~2.2.1-2.2.2) 
after including in eq.~(\ref{utot}) the energy density of the once EC scattered photons ($u_{\rm EC}$). 
The interested readers can compare how the inclusion of $u_{\rm EC}$ alters the results presented in Sect.~\ref{AR}.

At this point, it is useful to introduce the compactness of the magnetic and photon fields that appear in eq.~(\ref{utot}).
We define the compactness of a field, which is a dimensionless measure of its energy density, as
\eqb
\ell_{\rm j} = \frac{\sth \rb u_{\rm j}}{\mel c^2},
\label{comp}
\eqe
where the subscript $j$ takes the values `B' (magnetic field), `ex' (external photon field), `syn' (synchrotron photons)
and `ssc,i' ($i$-th generation  of SSC photons). 
Similarly, one can define the injection compactness of electrons, which is given by
\eqb
\linj  = \frac{\sth L_{\rm e}^{\rm inj}}{4\pi \rb \mel c^3} = \frac{\rb^2 \sth Q_0 \gamma_0}{3c}.
\label{leinj}
\eqe

In most astrophysical applications, 
the total energy density $u_{\rm tot}$ is dominated by the first two or three terms, namely by
the energy density of synchrotron photons and that of the magnetic field (or of ambient photon fields).
In such cases, analytical time-dependent solutions of eq.~(\ref{kinetic}) can be found (see \citealt{kardashev62, zacharias10, zacharias12}).
However, if  $\sum_{i=1}^{\nt-1} u_{\rm ssc,i}$ becomes comparable\footnote{We will quantify this statement in Sects.~2.2.1-2.2.2} to the sum of the first three terms, then
even the derivation of steady-state solutions of eq.~(\ref{kinetic}) becomes challenging, as the energy density of each SSC generation
depends on $\nel$ itself. {In such a case, one can still derive analytical steady-state solutions of eq.~(\ref{kinetic}), but 
only in two limiting regimes, depending on which is the dominant term in the kinetic equations. 

In order to define
the aforementioned regimes, we introduce the ratio  $\xi$
of the cooling and escape  timescales of electrons with Lorentz factor $\gamma_0$:
\eqb
\xi = \frac{3 \mel c^2}{4 \sth \rb u_{\rm tot} \gamma_0}= \frac{3}{4 \ell_{\rm tot}\gamma_0},
\label{cool_esc}
 \eqe
 where $\ell_{\rm tot}$ is defined accordingly to eq.~(\ref{comp}).

We distinguish between two regimes\footnote{We caution the reader that the slow and fast cooling regimes defined here
are, strictly speaking,  valid only asymptotically, i.e. for $\xi \gg 1$ and $\xi \ll 1$.}:

\begin{itemize}
 \item If $\xi > 1$, then electron cooling 
 is negligible and the steady-state electron distribution is determined
by the balance between the escape and injection processes. 
The steady-state electron distribution is then simply 
$\nel(\gamma)/\tesc = Q_{\rm e}\left(\gamma \right) $ or equivalently
$\nel(\gamma)= \ks\delta(\gamma-\gamma_0)$. Borrowing the terminology from GRBs,
we will refer to this regime as {\sl slow cooling} regime (see e.g. \citealt{sari98}).
\item If $\xi < 1$, then electron cooling becomes important. In this case, the cooling Lorentz factor
of electrons, which is defined by $\xi(\gc)=1$, 
is smaller than the injection one ($\gc < \gamma_0$) and the steady-state electron distribution
has the well-known $\gamma^{-2}$ power-law distribution (for mono-energetic injection), 
i.e. $\nel(\gamma) = \kf\gamma^{-2}$, for $\gc<\gamma \le \gamma_0$; this corresponds
to the so-called {\sl fast cooling} regime (e.g. \citealt{piran04}). 
\end{itemize}

The normalization factors $\ks$ and $\kf$, whose explicit form is presented in the following paragraphs, 
depend on the injection rate $Q_0$ and on the energy densities of the various fields.
In the paragraphs that follow, we will determine the slow and fast cooling regimes, and present
the respective analytical expressions for the compactnesses $\ell_{\rm j}$.
\subsubsection{Solutions in the slow cooling regime}
In the slow cooling regime and for an energy independent escape timescale,
the steady-state electron distribution has the same
energy dependence as the source function and the normalization $\ks$ depends
only on the injection rate as $\ks = Q_0\rb/c$.
Using the definition of eq.~(\ref{leinj})  we may write
\eqb
\ks = \frac{3 \linj}{\sth \rb \gamma_0}.
\label{ke-slow}
\eqe
The synchrotron and $i$-th generation SSC compactnesses are written as
\eqb
\label{lsyn-slow}
\lsyn & =  & 4 \lb \linj \gamma_0\\
\lssc & = & \lb \left(4\linj \gamma_0\right)^{i+1},
\label{lssc-slow}
\eqe
where we used eqs.~(\ref{ke-slow}), (\ref{us}) and (\ref{ussc}) (see Appendix~\ref{app1} for more
details). The above expressions
demonstrate the well-known linear and quadratic dependence of the synchrotron and first SSC components
on $\linj$ (e.g. \citealt{bloommarscher96}). Since $\lssc \propto \lsyn^{i+1}$, a cubic relation 
between the synchrotron and second SSC photon components is expected, which 
might be of interest for flaring events in high-energy emitting blazars \citep{aharonian07}.

For $\linj < 1/4\gamma_0$,  expressions eqs.~(\ref{lsyn-slow}) and (\ref{lssc-slow}) 
suggest that $\lssc \ll \lsyn < \lb$, and the total 
compactness of the source, in this case, is $\ell_{\rm tot} \approx \lb+\lex$.
The spectral energy distribution (SED) of the source will be synchrotron dominated if $\lb > \lex$ and EC dominated, otherwise.

The compactness of successive SSC photon generations will become progressively higher, namely the source
may enter the inverse Compton catastrophe regime, only if
$\linj > 1/4 \gamma_0$. In this case, the following ordering among the compactnesses holds:
\eqb
\ell_{\rm ssc, \nt-1} \gg \lsyn > \lb.
\eqe
The total compactness is, in a good approximation, given by
\eqb
\ell_{\rm tot} \approx \lex +  \lb \left(4\linj \gamma_0\right)^{\nt}.
\label{ltot-slow}
\eqe
Depending on the relative ratio of the two terms appearing above,
the SED is expected to be dominated either by the EC component or by the highest order SSC generation.
Thus, even for an external photon field with high compactness, the production of higher order, and progressively more luminous,
SSC photon generations cannot be suppressed for a sufficiently high electron compactness, 
namely $\linj > (1/4\gamma_0)(\lex/\lb)^{1/\nt}$. This will be demonstrated in more detail with numerical results in Sect.~\ref{NR}.

The results we derived so far are valid as long as the assumption of slow cooling is valid, and as such an {\sl a posteriori} check
is necessary. The expressions (\ref{ke-slow})-(\ref{ltot-slow}) are valid as long as $\xi > 1$ or equivalently 
$\ell_{\rm tot} < 3/4\gamma_0$, where $\ell_{\rm tot}$ is given by
\eqb
\ell_{\rm tot} \approx \left\{ \begin{array}{l l}
			      \lex +  \lb, & \linj < \frac{1}{4\gamma_0} \\
			      \phantom{} & \phantom{} \\
                             \lex +  \lb \left(4\linj \gamma_0\right)^{\nt}, & \linj > \frac{1}{4\gamma_0}                          .
                             \end{array}
                              \right.
                              \label{ltot_slow}
                              \eqe
By combining the above,             
we find that the slow cooling solutions are valid, if the parameters satisfy the following conditions:
\eqb
\label{cons1}
\linj & < & \frac{1}{4\gamma_0}  \ {\rm and} \\
f_{\rm ex} & < & \frac{3}{4\lb \gamma_0},
\label{cons2}
\eqe 
or 
\eqb
\label{cons3}
\linj & > & \frac{1}{4\gamma_0}  \ {\rm and} \\
f_{\rm ex} & < & \frac{3}{4\lb \gamma_0}+1-\left(4\linj \gamma_0 \right)^{\nt},
\label{cons4}
\eqe 
where 
\eqb
f_{\rm ex} \equiv 1 + \frac{\lex}{\lb}.
\label{fex}
\eqe
The quantity $f_{\rm ex}$ characterizes the relative strength of the external photon energy density
relative to the magnetic field energy density, and as such the importance of EC relative to synchrotron 
and SSC emission. It is one of the two parameters that we use to characterize the relevant parameter phase space. 
Relations (\ref{cons1})-(\ref{cons2}) and (\ref{cons3})-(\ref{cons4}) can be used to
define two distinct regions in the slow cooling parameter space. We will refer to them as $S_1$ and $S_2$, respectively.

\subsubsection{Solutions in the fast cooling regime}
\label{fastcool}
In the fast cooling regime the steady-state
equation of electrons is written as
\eqb
\frac{\partial}{\partial \gamma}\left(\gamma^2 \nel(\gamma)\left[\ub+\uex+\us+
\sum_{i=1}^{\nt-1} u_{\rm ssc,i}\right]\right) = \frac{3 Q_{\rm e}(\gamma) \mel c^2}{4\sth c}.
\eqe
An Ansatz for the solution $\nel$ of the above equation is $\nel=\kf\gamma^{-p}$. 
By substituting it in the above equation we
find  that $p=2$ and that $\kf$ satisfies the following algebraic equation
\eqb
\kf \left(f_{\rm ex} + \alpha \kf + \sum_{i=1}^{\nt-1} \left(\alpha \kf\right)^{i+1} \right) =  
\frac{9\linj/\lb}{4\sth \rb \gamma_0},
\label{general}
\eqe
where 
\eqb
\alpha \equiv 4\sth \rb \gamma_0/3.
\label{aa}
\eqe
Given that there is no general formula
for the roots of a polynomial with degree higher than four, which corresponds to $\nt=3$, 
explicit solutions of eq.~(\ref{general}) can be found in limiting cases only, which we list below.
\begin{itemize}
 \item {\sl Synchrotron and inverse Compton cooling (on synchrotron and external photons)}\\
  The main energy loss channels for electrons 
 are synchrotron cooling and inverse Compton scattering 
 on external and synchrotron photons,  i.e. $\lb+\lex+\lsyn \gg \sum_{i=1}^{\nt-1} \ell_{\rm ssc,i}$. In this case, $\kf$ 
 satisfies a second degree polynomial equation, which has the solution
 \eqb
\label{ke1-fast}
\kf = \frac{f_{\rm ex}}{2\alpha}\left(-1+\sqrt{1+ \frac{12 \linj/ \lb}{f_{\rm ex}^2}} \right).
\eqe
The above expression for $\lex \ll \lb$  simplifies into the one obtained for SSC electron cooling (see also \cite{petrolefa13}).
Using eqs.~(\ref{comp}) and (\ref{us}), we find  the synchrotron photon compactness to be
\eqb
\label{lsyn1-fast}
\lsyn = \lb \alpha \kf  = \lb\frac{f_{\rm ex}}{2}\left(-1+\sqrt{1+ \frac{12 \linj/ \lb}{f_{\rm ex}^2}} \right),
\eqe
while the compactness of  $i$-th SSC generation is simply 
\eqb
\lssc = \lb \left(\frac{\lsyn}{\lb}\right)^{i+1}.
\label{lssc1-fast}
\eqe
{ Expressions (\ref{lsyn1-fast}) and (\ref{lssc1-fast}) demonstrate how the inclusion of electron cooling
affects the dependence of the photon compactness of various components 
on the injected electron luminosity. For example, the  
synchrotron luminosity does not depend linearly on $\linj$ as in the slow cooling regime \citep[e.g.][]{bloommarscher96}, unless
$12 \linj/ \lb f_{\rm ex}^2 \ll 1$. }

We remind the reader 
that the expression for $\kf$ was derived 
under the assumption  of $\alpha \kf < 1$ (see also eq.~(\ref{general})) that allowed us to drop
the sum over the energy densities of higher-order SSC photons. The condition  $\alpha \kf < 1$ translates
into the following constraint
\eqb
f_{\rm ex} > 3\frac{\linj}{\lb}-1.
\label{conf1}
\eqe
\item  {\sl Cooling on higher order SSC photon generations}\\
Here we consider the case where { the system is driven in the Compton catastrophe regime and}
the dominant energy density is $\sum_{i=1}^{\nt-1} \ell_{\rm ssc,i} \approx \ell_{\rm ssc, \nt-1}$.
The energy density of external photons may dominate over that of SSC photons, unless
\eqb
\left(\alpha \kf \right)^{\nt} > f_{\rm ex}-1.
\label{conf2}
\eqe 
In this limit, 
the solution of eq.~(\ref{general}) is given by
 \eqb
\label{ke2-fast}
\kf  =  \frac{1}{\alpha} \left(\frac{3\linj}{\lb} \right)^{1/(\nt+1)}.
\eqe
The above expression is valid as long as the condition (\ref{conf2}) holds, or equivalently if 
\eqb
f_{\rm ex} < 1+\left(\frac{3\linj}{\lb}\right)^{\nt/(\nt+1)},
\label{conf3}
\eqe
where we substituted expression (\ref{ke2-fast}) into eq.~(\ref{conf2}). This is the second constraint that we derive
in the fast cooling regime.

Using the same equations from the Appendix~\ref{app1} as before, we find that 
\eqb
\label{lsyn2-fast}
\lsyn & = & \lb  \left(\frac{3\linj}{\lb} \right)^{1/(\nt+1)}\\
\lssc & = & \lb  \left(\frac{3\linj}{\lb} \right)^{(i+1)/(\nt+1)}.
\label{lssc2-fast}
\eqe
                             
\end{itemize}
Before closing this paragraph we complete
the set of constraints for the fast cooling regime.
For this, we require the condition  $\xi < 1$ or $\ell_{\rm tot} > 3/4\gamma_0$  to hold. Using 
the adequate expression for the total source compactness, i.e. 
\eqb
\ell_{\rm tot} \approx 
\left\{ \begin{array}{ll}                            
                             \lex +  \lb, & f_{\rm ex} > 3\frac{\linj}{\lb}-1 \\
                              \lex +\lb  \left(\frac{3\linj}{\lb} \right)^{\frac{\nt}{\nt+1}}, &  f_{\rm ex}  < 1+ 
                             \left(3\frac{\linj}{\lb}\right)^{\nt/(\nt+1)}
                              \end{array}
                              \right.
                                 \label{ltot_fast}
                              \eqe
we find the following constraints:
\eqb
\label{conf4}
f_{\rm ex} &  > & \max\left(\frac{3}{4\lb \gamma_0}, 3\frac{\linj}{\lb}-1 \right)
\eqe
or
\eqb
\label{conf5}
f_{\rm ex} & < & 1+\left(3\frac{\linj}{\lb}\right)^{\nt/(\nt+1)} \ {\rm and}\\
\linj & > & \frac{\lb}{3}\left(\frac{3}{4\lb\gamma_0} \right)^{\frac{\nt+1}{\nt}}.
\label{conf6}
\eqe
Similar to what we found for the slow cooling regime, the 
constraints (\ref{conf4}) and (\ref{conf5})-(\ref{conf6})
divide the fast cooling parameter regime into two sub-regions. We will refer
to them as regions $F_1$ and $F_2$, respectively.

We derived the conditions 
that result in the dominance of the inverse Compton emission over the synchrotron one, both in the slow
and fast cooling regime. All the conditions were expressed, so far,  in terms of the various compactnesses that describe the source.
The strength  of the inverse Compton emission is traditionally expressed also in terms of the Compton $Y$ parameter \citep{rybicki79}.
For completeness, we present the relation between the Compton $Y$ parameter and the
various compactnesses in Appendix~\ref{definitions}.  {The interested
reader can use this relation to express all the analytical expressions derived in Sect.~\ref{AR}
in terms of the $Y$ parameter.}

\subsubsection{The $f_{\rm ex}-\linj$ parameter space}
\label{parameterspace}
\begin{figure*}
 \centering
 \includegraphics[width=0.44\textwidth]{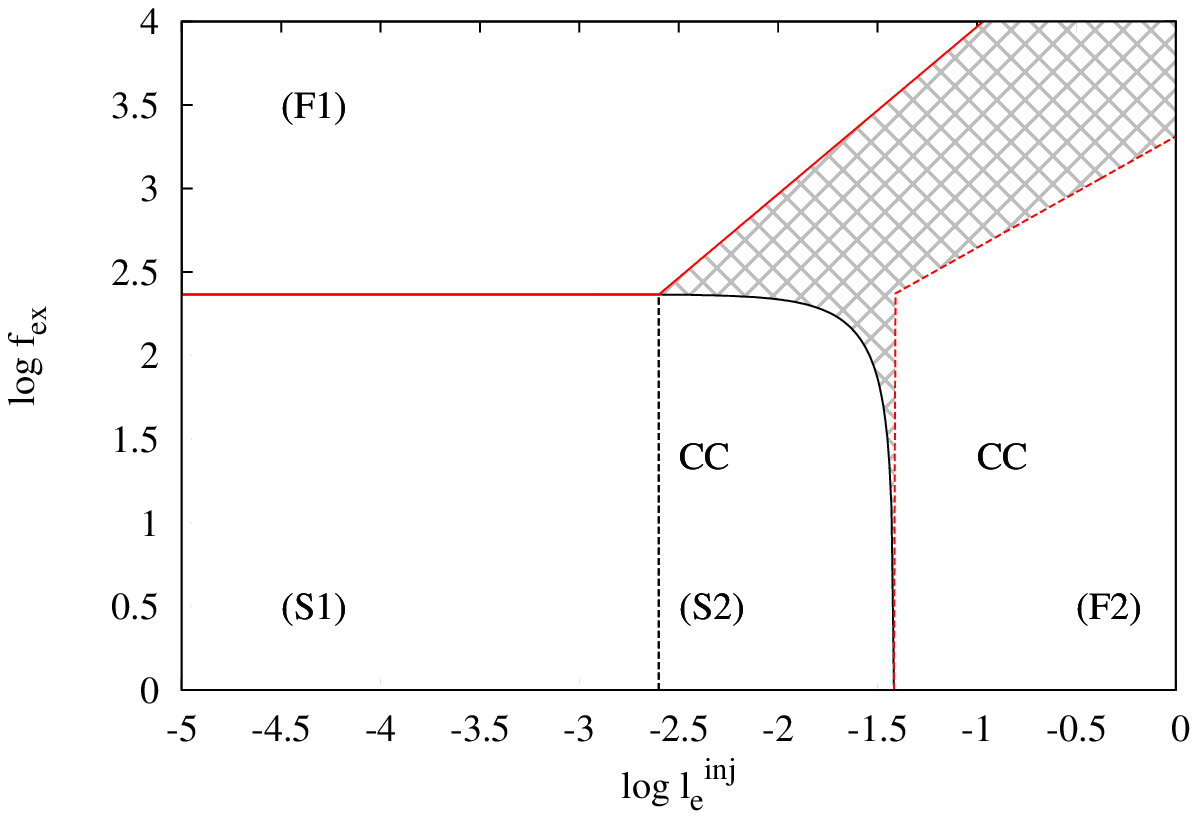}
  \includegraphics[width=0.44\textwidth]{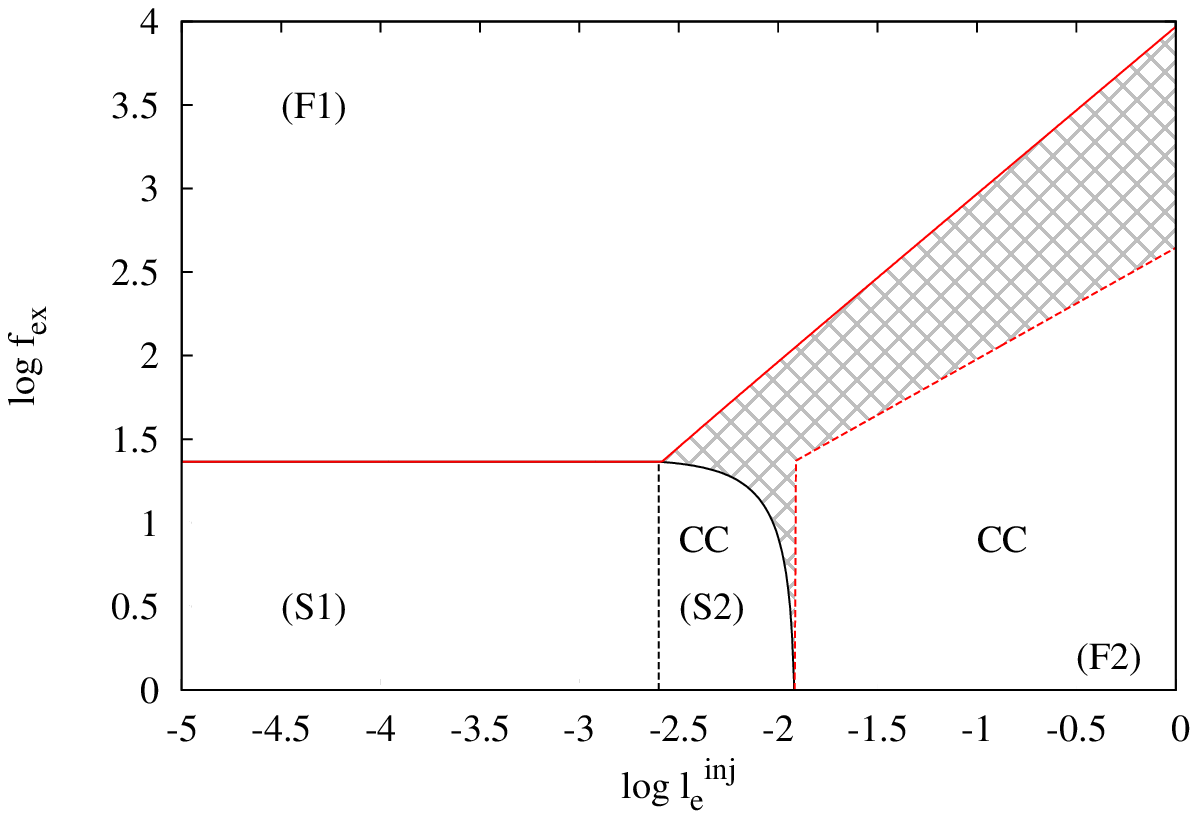}
  \caption{Log-log plot of $f_{\rm ex}$ versus $\linj$ for $\gamma_0=100$. Left and right panels 
  are obtained for $\lb=3\times 10^{-5}$ and $3\times 10^{-4}$, respectively.
  The constraints defined by eqs.~(\ref{cons1})-(\ref{cons2}) and (\ref{cons3})-(\ref{cons4})  are shown respectively
  with black dashed and solid lines. The constraint imposed by eq.~(\ref{conf4}) is plotted with a solid
  red line, while the red dashed line corresponds to the constraint given by eqs.~(\ref{conf5})-(\ref{conf6}). The various sub-regions
  defined in text are marked on the plot as $S_{1,2}$ and $F_{1,2}$. The gray-hatched region denotes
  the parameter regime in the fast cooling limit, where our analysis is not valid.}
  \label{par-space}
\end{figure*}
The regimes defined in the previous paragraphs are illustrated
in Fig.~\ref{par-space}, where we plot $f_{\rm ex}$ against $\linj$ for 
$\gamma_0=100$ and two values of the magnetic compactness, i.e. $\lb=3\times10^{-5}$ (left panel)
and $\lb=3\times 10^{-4}$ (right panel). 
For a given $\lb$, $f_{\rm ex}$  translates into a value of $\lex$, and thus, from 
the $f_{\rm ex}-\linj$ plot one can read the values
of all the basic compactnesses that describe the source. 
{For $\linj \le 1$ the source is found to be optically thin to Thomson scatterings,  
where our analysis is valid (see also Sect.~\ref{processes}).}

On both plots, there are some regions marked
as $S_{1,2}$ and $F_{1,2}$, which were respectively defined in 
Sections 2.2.1 and 2.2.2. To demonstrate the regions of the parameter space where the inverse 
Compton catastrophe occurs, we used the abbreviation \cc.
The dashed and solid black lines correspond to the slow cooling regime
constraints of eqs.~(\ref{cons1})-(\ref{cons2}) 
and (\ref{cons3})-(\ref{cons4}), respectively. The constraints obtained in the fast cooling regime, which 
are given by eqs.~(\ref{conf4}) and eqs.~(\ref{conf5})-(\ref{conf6}) are shown with solid and dashed red lines, respectively.
Finally, the transition region between the two limiting cases  in the fast cooling regime (see Sect.~2.2.2),
is not tractable  by analytical means.

The main conclusions drawn from Fig.~\ref{par-space} are summarized below:
\begin{itemize}
\vspace{-0.2cm}
 \item for a fixed $\lb$, the system may enter the fast cooling regime either for high values of $\linj$ (lower right region)
 or high external photon compactnesses (upper left  and middle regions). 
 In the region $F_1$, the main channel of electron energy losses is synchrotron radiation
 and EC scattering. In fact, there is a transition from synchrotron to EC scattering cooling, as one moves towards
  higher values of $f_{\rm ex}$ within the $F_1$ region.
 \item in the $F_2$ region, electrons lose energy preferentially
 by the SSC process. Moreover, the energy density of higher order SSC photon generations becomes
 larger, and as such, it dominates the total energy density which appears in the electron cooling
 term. 
 \item the regions $S_1$ and $S_2$ of the slow cooling regime
 occupy the lower left and middle part of the parameter
 space, where both the electron and external photon compactnesses are low/moderate.
 \item the inverse Compton catastrophe may occur either in the slow cooling or in the fast cooling
 regimes. Higher values of $\linj$ are required in the latter.
 \item even for $\lex \gg \lb$ (or $f_{\rm ex}\gg 1$) the inverse Compton catastrophe cannot be avoided, if the 
 compactness of electrons is sufficiently high (see also eqs.~(\ref{cons3}) and (\ref{conf6})).
 \item an increase of $\lb$ results in   a more extended fast cooling regime, as expected (see 
 e.g. regions $F_1$ in the left and right panels of Fig.~\ref{par-space}).
 \item the minimum $\linj$, which is required for the inverse Compton catastrophe to occur while in the fast cooling regime, decreases
 as $\propto \lb^{-1/\nt}$. 
\end{itemize}

\subsection{Observed luminosities}
\label{luminosities}
Having investigated the different regimes of the available parameter space, we proceed  with
the calculation of observed quantities, and in particular,
of the bolometric synchrotron and SSC luminosity. 
For a spherical region moving relativistically 
with a Doppler factor $\Dop = \Gamma^{-1} \left(1-\beta \cos \thobs \right)^{-1}$, where
$\Gamma$ is the bulk Lorentz factor and $\thobs$ is the angle between the beaming direction
 and the line of sight to the observer, the observed synchrotron and SSC luminosities
are easily derived by the respective compactnesses as:
\eqb
L_{\rm i}^{\rm obs} = \ell_{\rm i}\frac{4 \pi \rb \mel c^3 \Dop^4 }{\sth}, \ i={\rm syn, ssc},
\label{bolometric}
\eqe 
where $\ell_{\rm i}$ is given by eqs.~(\ref{lsyn-slow})-(\ref{lssc-slow}) 
for the slow cooling regime and by eqs.~(\ref{lsyn1-fast})-(\ref{lssc1-fast}) and (\ref{lsyn2-fast})-(\ref{lssc2-fast})
for the fast cooling regime. From this point on, all 
quantities noted with the subscript `obs' will refer to the observer's frame, while quantities with an asterisk ($\star$) 
will be measured in the rest frame of the central engine/galaxy.
\begin{figure*}
 \centering
 \includegraphics[width=0.8\textwidth]{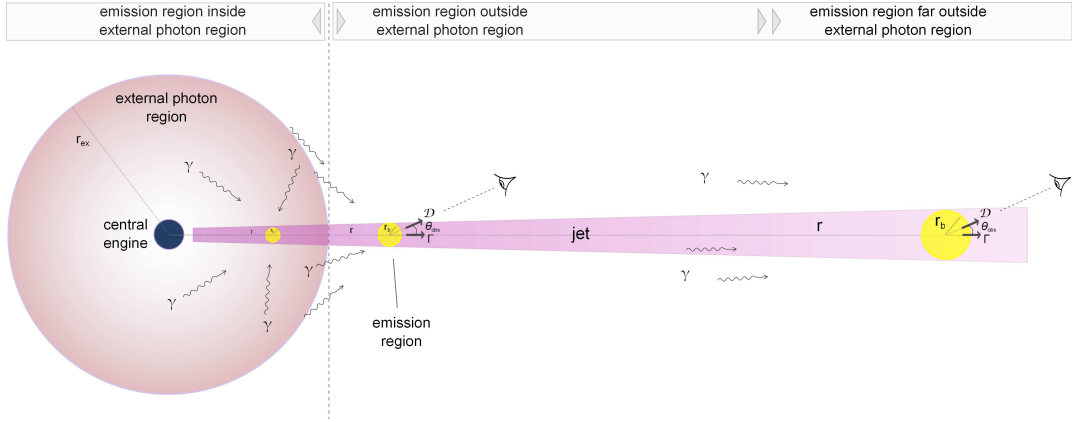}
 \caption{Illustration of the physical system under consideration. The emission region (yellow blob) 
 is part of a relativistic jet launched by the central engine, and  lies at a distance $r$ from it. 
 In the rest frame of the central engine, the external photon field is isotropic and spherical with radius $\rex$ (purple region).
 If $r \lesssim \rex$, the energy density of the external photon field is significantly boosted ($\propto \Gamma^2 u_{\rm ex, \star}$) because of 
 the relativistic motion of the emission region. The opposite holds for cases where $r \gg  \rex$, while the effect
 of Lorentz boosting is intermediate for $r\gtrsim \rex$.}
 
 
 \label{sketch}
\end{figure*}

The details about the external source of photons, such as its geometry and spectral energy distribution, 
were not important for the derivation of the results presented in Sect.~2.2. Before stepping into the calculation
of the EC observed luminosity, it  is, however, useful to specify the so far arbitrary source of external photons. 
We  consider the case of an isotropic (in the galaxy rest frame)}, monochromatic 
 photon field with a photon energy $\eex$ and  spherical geometry with a radius 
 $\rex$ (see Fig.~\ref{sketch}). Its energy density is  $u_{\rm ex, \star} =\Lex/ 4\pi \rex^2 c$, where $\Lex$ is its luminosity.
 If the emission region is embedded in the external photon field, 
 the energy density and the characteristic photon energy of this field in the comoving frame of the emission region appear  as 
 $\uex \simeq \Gamma^2 u_{\rm ex, \star}$ and $\epsilon_{\rm ex}\simeq \Gamma \eex$, respectively \citep{dermerschlickeiser93, dermer95}.
 If, however, the emission region is located at a distance $r \gg \rex$ 
 and is moving away from it with a Lorentz factor $\Gamma$ such that $r /\rex \gg  \Gamma$ (see e.g. Fig.~\ref{sketch}), 
 the external energy density will be then significantly (by a factor of $\sim1 /\Gamma^2$) de-boosted (e.g. \citealt{dermer95}). In this
 case, the system can be satisfactorily studied by taking into account only the internally produced photons.
 Finally, for cases where $r /\rex\gtrsim \Gamma$ the boosting effect on the energy density and typical photon energies
 of the external radiation field will be intermediate.

For the observed EC luminosity ($L_{\rm EC}^{\rm obs}$)  we follow the approach by \cite{dermer95} -- henceforth D95, since
the external photon field in the comoving frame of the emission region
is not isotropic. For the slow cooling regime, where the stationary electron distribution is
mono-energetic, i.e. $\nel = \ks \delta(\gamma-\gamma_0)$, we start with eq.~(5) of D95 and
derive (see Appendix~\ref{app2} for more details) the observed flux due to EC scattering 
\eqb
F_{\rm EC}^{\rm obs}(\eobs, \Omega_{\rm obs})  = F_0 
\left(\frac{\Dop  (1+\mobs)}{1+\beta}\right)^{3/2} \!\!\left(\frac{\eobs(1+z)}{\Dop \eex} \right)^{1/2}\!\!
\delta \left(\eobs - \epsilon_{\rm EC} \right),
\label{Fc}
\eqe
where $\beta=\sqrt{1-1/\Gamma^2}$, $\epsilon_{\rm EC}$ is defined as 
\eqb
  \epsilon_{\rm EC} \equiv  \frac{\Dop^2 \eex \gamma_0^2 (1+\mobs)}{(1+z)(1+\beta)},
 \label{eec}
\eqe
where $\mobs=\cos\theta_{\rm obs}$ and $F_0$ is given by
\eqb
\label{F0}
F_0  = \frac{\Dop^4 \vb \sth c}{4\pi D_{\rm L}^2} u_{\rm ex} \ks \gamma_0.
\eqe
After integrating eq.~(\ref{Fc}) over $\eobs$ and making the approximations $\thobs \simeq 1/\Gamma$,
$\mobs \approx 1- 1/2 \Gamma^2$, $1+\beta \simeq 2$ and  $\Dop \simeq \Gamma$ we find
\eqb
L_{\rm EC}^{\rm obs}  \approx  \Gamma^6 \vb \sth c u_{\rm ex} \ks \gamma_0^2.
\label{Lec-1}
\eqe
For the fast cooling regime, where the stationary electron distribution is a power-law, 
we can use directly eq.~(7) in D95. After making the  substitutions
$K \rightarrow \kf$ and $\alpha=1/2$, where $\alpha=(p-1)/2$, we integrate  
 eq.~(7) over energies up to $\eobs \approx \Dop^2 (1+\mobs)\gamma_0^2 \eex / (1+\beta) (1+z)$, and by
making the same approximations as before, we derive a similar expression to that of eq.~(\ref{Lec-1}):
\eqb
L_{\rm EC}^{\rm obs} \approx \Gamma^6 \vb \sth c u_{\rm ex} \kf \gamma_0.
\label{Lec-2}
\eqe
Thus, $L_{\rm EC}^{\rm obs}$ depends on the cooling regime only through the factors $\ks$ and $\kf$.

For the expressions regarding the slow cooling regime we made use of 
eqs.~(\ref{ke-slow})-(\ref{lssc-slow}), (\ref{bolometric}), and (\ref{Lec-1}), whereas for the fast cooling regime we used
eqs.~(\ref{ke1-fast})-(\ref{lssc1-fast}), 
(\ref{lsyn2-fast})-(\ref{lssc2-fast}), (\ref{bolometric}), and (\ref{Lec-2}). Our results are summarized below. 
\begin{itemize}
\item[-]{\sl Slow cooling} regime\\
\eqb
\label{lum-syn-slow}
L_{\rm syn}^{\rm obs} & = & \frac{3c}{\rb} \Gamma^4 \vb \ub \left(4\linj \gamma_0\right) \\
\label{lum-ssc-slow}
L_{\rm ssc, i}^{\rm obs} & = & \frac{3 c}{\rb} \Gamma^4 \vb \ub \left(\linj 4 \gamma_0\right)^{i+1} \\ 
L_{\rm EC}^{\rm obs} & = &\frac{3c}{4 \rb}\Gamma^6 \vb u_{\rm ex}  \left(4\linj \gamma_0\right).
\label{lum-ec-slow}
\eqe
\item[-]{\sl Fast cooling regime} \\
In Section \ref{fastcool} we derived explicit expressions for the compactness of various components for two  limiting cases:
\begin{itemize}
\item[$\bullet$] Synchrotron and inverse Compton cooling on synchrotron and external photons \\
\eqb
L_{\rm syn}^{\rm obs} & = & \frac{3 c}{2 \rb} \Gamma^4 \vb \ub f_{\rm ex}\left(-1+\sqrt{1+ \frac{12 \linj/ \lb}{f_{\rm ex}^2}} \right) \label{fast1-Lsyn}\\
L_{\rm ssc, i}^{\rm obs} & = & \frac{3 c}{2^{i+1}\rb} \Gamma^4 \vb \ub \left[f_{\rm ex}\left(-1+\sqrt{1+ \frac{12 \linj/ \lb}{f_{\rm ex}^2}} \right)\right]^{i+1} \label{fast1-Lssc}\\ 
L_{\rm EC}^{\rm obs} & = &\frac{3c}{8\rb}\Gamma^6 \vb u_{\rm ex} f_{\rm ex}\left(-1+\sqrt{1+ \frac{12 \linj/ \lb}{f_{\rm ex}^2}} \right).
\label{fast1-Lec}
\eqe
\item[$\bullet$] Cooling on higher order SSC photon generations (CC limit)\\
\eqb
L_{\rm syn}^{\rm obs} & = & \frac{c}{3\rb} \Gamma^4 \vb \ub \left(\frac{3\linj}{\lb} \right)^{1/(\nt+1)}\label{fast2-Lsyn} \\
L_{\rm ssc, i}^{\rm obs} & = & \frac{c}{3 \rb} \Gamma^4 \vb \ub \left(\frac{3\linj}{\lb} \right)^{(i+1)/(\nt+1)} \label{fast2-Lssc} \\ 
L_{\rm EC}^{\rm obs} & = & \frac{3c}{4\rb} \Gamma^6 \vb u_{\rm ex} \left(\frac{3\linj}{\lb} \right)^{1/(\nt+1)}.
\label{fast2-Lec}
\eqe
\end{itemize}
\end{itemize}
We remind  the reader that the above expressions are strictly valid
in the slow (regions $S_{1,2}$) and fast (regions $F_{1,2}$) cooling regimes, and one should use them with caution 
for parameters falling in the transition regime (e.g. hatched regions in Fig.~\ref{par-space}).

Thanks to the general formalism we adopted, i.e. 
all the observed luminosities are expressed in terms of the electron, magnetic and external radiation compactnesses, 
eqs.~(\ref{lum-syn-slow})-(\ref{fast2-Lec}) can  easily be applied to a variety of astrophysical sources, such as $\gamma$-ray emitting
blazars and GRBs.

\subsection{Photon spectrum in the CC limit}
Besides the observed luminosity of the various components that
build the multi-wavelength spectrum in the Compton catastrophe regime,
we are also interested in the spectral shape.
As already described in Sect.~\ref{intro}, the inverse Compton catastrophe 
refers to the case where higher order SSC photon generations 
are progressively more luminous, i.e. 
$L_{\rm ssc, i}^{\rm obs} > L_{\rm ssc, i-1}^{\rm obs}> \dots > L_{\rm syn}^{\rm obs}$.
The ordering of the emitting components  in terms of their luminosities, with the synchrotron component 
being the least luminous, implies that the photon spectrum
will be a power-law, defined by the peaks of successive components, which
will extend up to an energy $\sim 5(\gamma_0/10)$~MeV 
(as measured in the comoving frame).

Being equipped with all the necessary analytical expressions, we can turn now to the spectral index, $\beta_{\rm cc}$, 
(approximating   the photon spectrum  as a power-law, i.e. $L_{\epsilon}^{\rm obs} \propto \epsilon^{-\beta_{\rm cc}}$)
in the Compton catastrophe limit. 
Assuming that the bolometric luminosity of each component is a good
proxy for the peak luminosity, we can estimate $\beta_{\rm cc}$ through the approximate relation
\eqb
\frac{L_{\rm ssc, i+1}^{\rm obs}}{L_{\rm ssc, i}^{\rm obs}} \approx 
\left(\frac{\epsilon_{\rm ssc, i+1}}{ \epsilon_{\rm ssc,i}}\right)^{1-\beta_{\rm cc}}
\eqe
where  $\epsilon_{\rm ssc, i}\simeq (4/3)^{i}b \gamma_0^{2(i+1)}\mel c^2$. Using eqs.~(\ref{lum-ssc-slow}) and (\ref{fast2-Lssc}) for the slow
and fast cooling regimes, respectively, we find 
\eqb
\beta_{\rm cc} \simeq \left \{ \begin{array}{ll}
1-\frac{\log \left( 4\linj \gamma_0\right)}{\log \left( 4\gamma_0^2/3\right)}, & {\rm slow \ cooling }\\ 
\phantom{} & \phantom{} \\
1-\frac{\log \left(3 \linj/\lb \right)}{(\nt+1)\log \left( 4\gamma_0^2/3\right)}, & {\rm fast \ cooling}                                
                               \end{array}
\right.
\label{bcc}
\eqe
Equation~(\ref{bcc})  reveals the dependence of the spectral slope
 on various physical parameters. For example, in the fast cooling regime, a
decrease of the magnetic field strength and/or of
the size of the emission region, results in lower $\lb$ values, and thus, tends to make the photon spectrum harder (smaller $\beta_{\rm cc}$ values). 
{Notice that $\beta_{\rm cc}$ is independent of the magnetic compactness in the slow cooling regime, while 
 higher $\linj$ lead to harder photon spectra in both regimes.}
In principle, there is no lower limit on $\beta_{\rm cc}$. However, very hard photon spectra
with $\beta_{\rm cc}\lesssim 0.5$ would require unphysically large $\linj$ {due to 
the logarithmic dependence on the latter. In any case,
our results cannot be extended to $\linj \gg 1$, since in this parameter regime other processes, such as 
photon-photon absorption and pair cascades, become important and they cannot be neglected anymore.}

\begin{figure}
\centering
 \includegraphics[width=0.5\textwidth]{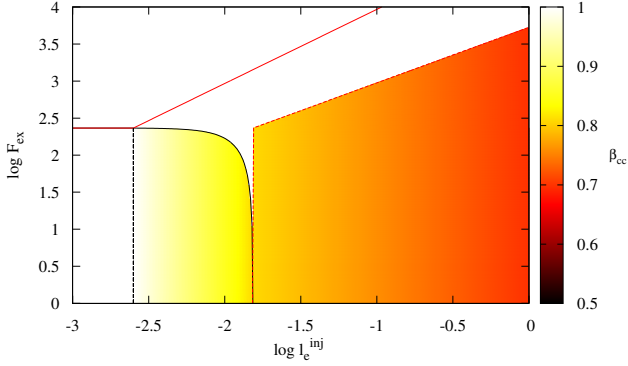}
 \caption{The $f_{\rm ex}-\linj$ parameter space for $\lb=3\times10^{-5}$ and $\gamma_0=100$.
 The color coding denotes the values of 
the spectral index in the Compton catastrophe limit ($\beta_{\rm cc}$). 
}
 \label{beta}
\end{figure}

The spectral index given by eq.~(\ref{bcc}) is obtained for parameters
drawn from the sub-regions $S_2$ and $F_2$ in the $f_{\rm ex}-\linj$ parameter space.
This is exemplified in Fig.~\ref{beta}, where we adopted $\gamma_0=100$ and $\lb=3\times10^{-5}$.
In the fast cooling regime, the 
Compton catastrophe photon spectra tend to be harder than these of the slow cooling regime. Figure \ref{beta}
demonstrates also that the spectra in the Compton
catastrophe limit become harder as $\linj$ increases. However, because of the weak logarithmic dependence on $\linj$ 
(see color bar), very high values of $\linj$ are required for $\beta_{\rm cc} \lesssim 0.5$. 
Finally, we note that a choice of a lower $\gamma_0$ would not have 
a strong impact on $\beta_{\rm cc}$ (see eq.~(\ref{bcc})), but it might affect the extend of the sub-regions $S_2$ and $F_2$.

\section{Numerical results}
\label{NR}
The results presented in the previous section
were derived by solving the simplest form of the electron kinetic equation (see eq.~(\ref{kinetic}))
and by using a series of approximations. In this section, we present 
the results of a numerical treatment of the problem, which does not suffer from the simplifying assumptions  
used in Sect.~\ref{AR}, {albeit its own limitations.} Our aim is to test whether or not these 
simplifications have a radical effect on the results
regarding the transition to the Compton catastrophe regime, as well as the multi-wavelength spectral properties in this limit.

The numerical treatment of the problem allows us to 
augment the {relativistic} electron kinetic equation with more processes, such as pair injection due to photon-photon absorption, 
and to make use of the full expressions for the synchrotron and Compton emissivities {in the relativistic limit}. In addition, 
we write an accompanying equation for photons, which is coupled to the one for electrons.
The physical system is then described by
\eqb
\label{ne1}
\frac{\partial \nel}{\partial t} + \frac{\nel}{\tesc} & = & \mathcal{L}_{\rm e}^{\rm syn} + \mathcal{L}_{\rm e}^{\rm ics} + \mathcal{Q}_{\rm e}^{\gamma \gamma} +\mathcal{Q}_{\rm e} \\
\frac{\partial \nph}{\partial t} + \frac{\nph}{t_{\gamma,\rm esc}} & = & 
\mathcal{Q}_{\gamma}^{\rm syn} + \mathcal{Q}_{\gamma}^{\rm ics} + \mathcal{L}_{\gamma}^{\rm ssa} + \mathcal{L}_{\gamma}^{\rm \gamma \gamma}.
\label{ng1}
\eqe
The various terms appearing above describe the following physical processes: (i) Photon-photon pair production, 
which acts as a source {of new particles} ($\mathcal{Q}_{\rm e}^{\gamma \gamma}$)
and a sink term for photons ($\mathcal{L}_{\gamma}^{\rm \gamma \gamma}$); (ii) synchrotron radiation, which acts as  
{an energy} loss term for electrons ($\mathcal{L}_{\rm e}^{\rm syn}$) 
and a source term for photons ($\mathcal{Q}_{\gamma}^{\rm syn}$); (iii) synchrotron self-absorption, which acts as a 
{sink term}
for
photons ($\mathcal{L}_{\gamma}^{\rm ssa}$); and (iv) inverse Compton scattering, which
acts as {an energy} loss term for electrons\footnote{Both external and internal photons serve as targets for the inverse Compton
scattering process.} ($\mathcal{L}_{\rm e}^{\rm ics}$) and an {energy injection} term for photons ($\mathcal{Q}_{\gamma}^{\rm ics}$)
The escape timescale for electrons $\tesc$ is set equal to the crossing time of the source, i.e. $\tcr=\rb/c$.
Note that the expressions for the synchrotron and Compton emissivities  {in the
relativistic limit} (c.f. eqs.~(6.33) and (2.48) in \cite{rybicki79} and \cite{bg70}, respectively)
have been used in eq.~(\ref{ng1}).
For more details about the functional forms of the various rates, {see Appendix~\ref{app-num}.}

{Synchrotron self-absorption is taken into account only as a sink 
for photons, while there should be an additional source (heating) term for the electrons. Although
 this is not, strictly speaking, an energy conserving scheme,
in all cases we present the absorbed energy 
by synchrotron self-absorption is small, and electron heating can be neglected. Equations (\ref{ne1}) and (\ref{ng1}) could not
be applied in their present form in the ``synchrotron boiler''  case, where $\lb \gg \linj$ \citep{ghisellini88}.
}

{As the numerical code makes no distinction between electrons and positrons (both determined by eq.~(\ref{ne1})),
the treatment of the pair annihilation in the code is approximate \citep[see e.g.][]{coppi92}. 
As long as we consider optically thin cases, i.e. $\linj \lesssim 1-3$ (or, equivalently $\tau_{\rm e}< 1$ where $\tau_{\rm e}$ is
the Thomson optical depth and is defined in Appendix C)
the pair yield is very low and the annihilation can safely be neglected.
The annihilation line starts becoming important for $\linj>10$ \citep[e.g.][]{LZ87}, where
our numerical approach becomes invalidated.}

{In summary, the numerical code used in this paper and 
presented originally in \cite{mastkirk95} is adequate for the description 
of system containing relativistic electrons ($\gamma \gtrsim 3$) with compactnesses up to $\sim 3$.
Exploration of a system containing electrons with $\beta \gamma \lesssim 1$ and/or with very high compactness
requires more sophisticated numerical treatment \citep[e.g.][]{coppi92, sternetal95, belmont08, vurmpoutanen09}; this, however, lies out
the scope of the present study.}
\subsection{Comparison with analytical results}
As a first step, we compare the analytically derived values for the observed 
luminosities (see Sect. \ref{luminosities}) against the values
we obtained by numerically solving eqs.~(\ref{ne1}) and (\ref{ng1}), without including
the terms of synchrotron self-absorption and  
photon-photon absorption. In this way, we ensure that any differences found between
the numerical and analytical results will have to be related with the simplifying 
approximations of the analytical treatment and not 
with the physical processes themselves.

For the comparison we adopted the fiducial parameter set of $\gamma_0=100$, $\linj=10^{-2}$, $B=10$~G, $\rb=10^{16}$~cm, $\Gamma=10$.
Substitution of the $\gamma_0$ and $B$ values in eq.~(\ref{NT}) results in  $\nt=2$. 
The magnetic compactness is found to be $\lb=3\times 10^{-2} \gtrsim \linj$.
To test the effect of the external photon field on electron cooling and on the multi-wavelength photon spectra we 
used three indicative values for its compactness, namely $\lex=0, 0.1$ and $0.4$, which
correspond to equally spaced values of $\log f_{\rm ex}$. 
In all cases but $\lex=0$, the basic compactnesses are ordered as $\linj \lesssim \lb < \lex$.
{For the characteristic energy of external photons, we used the fiducial value $\eex=1$~eV.}

\begin{figure}
 \centering
 \includegraphics[width=0.44\textwidth]{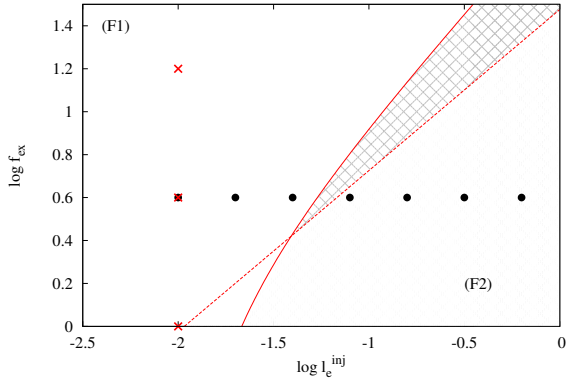}
  \caption{The $f_{\rm ex}-\linj$ plane for $\gamma_0=100$, $B=10$~G and $\rb=10^{16}$~cm ($\lb=3\times10^{-2}$).
 The parameter values used in the numerical runs of Sect.~\ref{NR} are shown as symbols:
 red crosses for the comparison between the analytical and numerical treatments (for more details, see text) and
 black circles for the transition of the system to the Compton catastrophe regime.
 The red lines and the grey-colored hatched region have the same meaning as in Fig.~\ref{par-space}. The sub-regions of
 the fast cooling regime, $F_1$ and $F_2$, are also marked on the plot.}
  \label{par-space-num}
\end{figure}

Using the analytical expressions of eqs.~(\ref{cons1})-(\ref{cons4}) and (\ref{conf4})-(\ref{conf6}) we construct
the respective  $f_{\rm ex}-\linj$ parameter space, which is illustrated in Fig.~\ref{par-space-num}.
The parameter values adopted for the comparison are shown as red crosses. For all three values of $\lex$, we analytically predict that 
the system should lie in the fast cooling regime (sub-region $F_1$). 
We note that even for $\lex=0$, the magnetic compactness ($\lb \propto B^2 \rb$) is high enough as 
to make the injected electrons cool efficiently.  This is indicated by the 
absence of the slow cooling regions in Fig.~\ref{par-space-num}. In what follows, we will compare
the aforementioned predictions with the respective numerical results.

\begin{figure}
\centering
\includegraphics[width=0.45\textwidth]{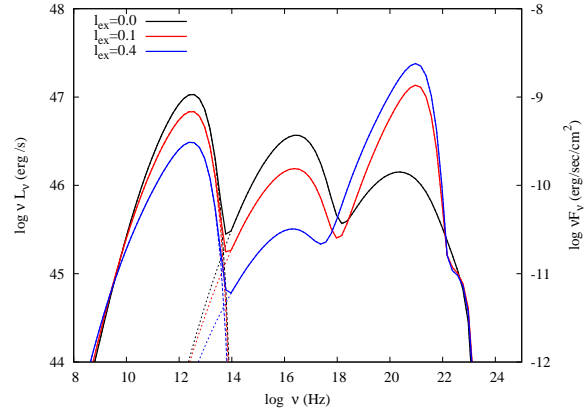} 
\caption{Steady-state multi-wavelength photon spectra calculated numerically for three values of $\lex$ marked on the plot.
The total emission is shown with thick solid lines, while the synchrotron and inverse Compton (SSC+EC) components are shown
with thin dashed and dotted lines, respectively.  Other parameters used are 
$\gamma_0=100$, $B=10$~G, $\rb=10^{16}$~cm, $\lb=3\times 10^{-2}$, $\linj=10^{-2}$, $\Gamma=10$,
$\eex=1$~eV. The flux values correspond to a source with a fiducial redshift $z=0.2$.}
\label{fig1}
\end{figure}

First, we verified that the steady-state electron distribution is described
by $n_{\rm e}(\gamma) \propto \gamma^{-2}$ as a result of efficient cooling,
for all values of $\lex$. The multi-wavelength photon spectra calculated numerically
for the three values of $\lex$ are shown in Fig.~\ref{fig1}.
Photon spectra obtained for $\lex=0, 0.1$ and $0.4$ are  plotted with black, red and blue lines, respectively. 
The synchrotron and inverse Compton components of the total  emission (thick line)
are plotted with dashed and dotted lines, respectively. 
For illustration reasons, the external radiation field is not shown. The flux values marked on the right vertical axis
correspond to a source placed at redshift $z=0.2$. 

In agreement with our analysis, we find two SSC photon generations being produced in the Thomson regime. This
is evident in the case of pure SSC emission (black lines). The increase of  $\lex$ has 
two effects on the shape of the SED. On the one hand, the synchrotron luminosity decreases, while, on the other hand,
there is a transfer of power from the SSC to the EC component (see e.g. red and blue lines in Fig.~\ref{fig1}).
This reflects the transition from synchrotron cooling to cooling due to EC scattering, which is in agreement with 
the analytical predictions (see Sect.~\ref{parameterspace}).

Apart from a qualitative comparison, we can compare the bolometric synchrotron, SSC and EC luminosities
as derived numerically and analytically from the appropriate expressions in Sect.~\ref{AR}, namely using eqs.~(\ref{fast1-Lsyn})-(\ref{fast1-Lec}).
The results are listed in Table~\ref{tab-1}. 
In all cases, the analytically derived values do not differ more than a factor of $\sim 2.5$
from those derived numerically. Given the simplifying assumptions we used to derive the analytical expressions in the first place,
we argue that the results are in good agreement to each other. 

\begin{table*}
 \caption{Observed bolometric luminosities of the synchrotron, SSC and external Compton components of the SED shown
 in Fig.~\ref{fig1}.}
 \begin{threeparttable}
 \begin{tabular}{c cc  cc  cc}
  \hline 
		& \multicolumn{2}{c}{$\lex=0$} & \multicolumn{2}{c}{$\lex=0.1$} &  \multicolumn{2}{c}{$\lex=0.4$}\\
  \hline
		& A\tnote{(a)}  & N\tnote{(b)} &  A & N & A & N \\       
\hline		
  $L_{\rm syn}^{\rm obs}$ (erg/s) &   $8.8\times10^{47}$ &  $4.0\times 10^{47}$ & $3.2\times10^{47}$	& $2.5\times10^{47}$	& $1.1\times 10^{47}$	& $1.2\times10^{47}$\\ 
  & & & & & \\
  $L_{\rm ssc,1}^{\rm obs}$ (erg/s) &  $5.1\times10^{47}$   &   $2\times 10^{47}$  &  $7\times10^{46}$ 	&  $7.9\times10^{46}$	& $7.1\times10^{45}$	& $1.5\times10^{46}$	\\ 
    & & & & & \\
  $L_{\rm ssc,2}^{\rm obs}$ (erg/s) &  $3.0\times10^{47}$   &  $6.3\times10^{46}$   &   $1.5\times10^{46}$	& hidden\tnote{(c)}     & $4.9\times10^{44}$	& hidden	\\   
    & & & & & \\
  $L_{\rm EC}^{\rm obs}$ (erg/s) &   -- & --    &   $2.5\times10^{47}$ &	  $4.7\times10^{47}$  & $3.2\times10^{47}$	& $7.9\times10^{47}$	\\     
    & & & & & \\
  \hline
 \end{tabular} 
  \begin{tablenotes}
  \item[a] Calculated using the analytical expressions given by 
          eqs.~(\ref{fast1-Lsyn})-(\ref{fast1-Lec}), which are valid in the fast cooling regime ($F_1$).
 \item[b] Calculated numerically -- see also Fig.~\ref{fig1}. 
 \item[c] The numerical code cannot distinguish between photons with EC and SSC origin, if they
 are produced at the same energy range. For example, if $L^{\rm obs}_{\rm ssc, 2} < L^{\rm obs}_{\rm EC}$, the 2nd SSC photon generation
 will be hidden by the EC one. Thus, we cannot numerically calculate its luminosity.
 
 \end{tablenotes}
 \end{threeparttable} 
 \label{tab-1}
  \end{table*} 
\subsection{Transition to the inverse Compton catastrophe}
In this paragraph we demonstrate the modification of the multi-wavelength photon spectra
caused by a gradual increase of the electron injection compactness, which 
eventually leads the source to the Compton catastrophe regime. 
Same as in previous paragraph, we neglect the effects of additional processes, such as
synchrotron self-absorption, which will be discussed separately in a following paragraph. 

We use as a starting point,  the fiducial parameter set of the previous section 
with $\lex=0.1$ (see Fig.~\ref{par-space-num}). In this case,
the external  photon energy density is the dominant term in the electron cooling. This
is also reflected to the SED
shown in Fig.~\ref{fig1}, where most of the bolometric photon luminosity is carried by the EC component.
We will show next that even in this case, higher order SSC photon generations may have a dominant contribution
to the SED and electron cooling, for sufficiently high $\linj$, as already pointed out in Sect. \ref{AR}.
Since this particular example falls in the fast cooling regime, we can use
the expressions for $L_{\rm ssc, \nt}$ and $L_{\rm EC}$ given by eqs.~(\ref{fast2-Lssc}) and (\ref{fast2-Lec}), respectively,
in order to predict the  necessary electron compactness for the highest SSC component to be the most luminous.
By demanding $L_{\rm ssc, 2} \ge L_{\rm EC}$, we find that $\linj \gtrsim 0.2$, {yet $\linj < 1$.}

We initiate the numerical calculations with $\linj=10^{-2}$ and keep increasing $\linj$
by a factor of two over its previous value until we reach $\linj=10^{-0.2}$. The predicted value for the transition to
the inverse Compton catastrophe is  slightly below the maximum value used in the simulation ($-0.7$ in logarithm).
The successive values are shown as black circles in the parameter space of Fig.~\ref{par-space-num}.
\begin{figure}
\centering
\includegraphics[width=0.45\textwidth]{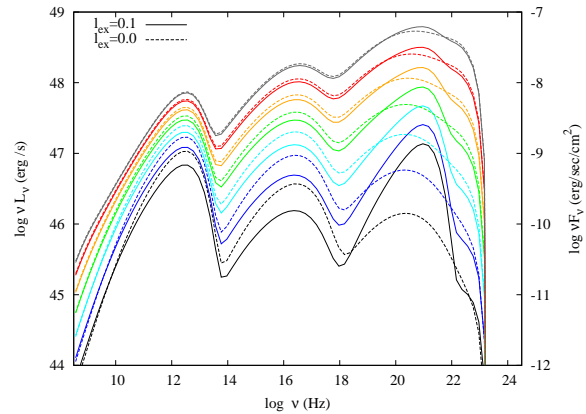} 
\caption{Steady-state multi-wavelength spectra obtained for different values of $\linj$, 
starting from $10^{-2}$ (black lines) and increasing up to $10^{-0.2}$ (grey lines)
with logarithmic increments of $0.3$. The multi-wavelength spectra obtained for $\lex=0$ are over-plotted with dashed lines, 
for comparison reasons. All other parameters
are the same as in Fig.~\ref{fig1}.}
\label{fig2}
\end{figure}

The successive photon spectra are shown in Fig.~\ref{fig2} with solid lines, while 
spectra {obtained for the pure SSC case, i.e. with $\lex=0$, are shown}
with dashed lines 
for comparison reasons. For the particular choice of parameters (see also Fig.~\ref{fig1}), we
find that the EC emission dominates in the $\gamma$-rays, while the SSC emission is being
suppressed, for low enough values of the electron compactness, i.e. $\linj=0.01-0.08$.
As the electron compactness progressively increases, however, the contribution of the EC component to the high-energy part of the
spectrum becomes  less significant.
We find, in particular, that it is the second SSC emission that actually 
dominates the multi-wavelength emission for $\linj\ge 10^{-0.8}$ (second spectrum from the top).
Note that this value  
is in good agreement with the analytical estimate derived above.
Finally, the spectrum obtained for $\linj=10^{-0.2}$ is a typical example of
emission in the Compton catastrophe regime {\citep[see also Fig.~2 in][]{bjornsson00}} and bears several `generic' features:
\begin{enumerate}
 \item The bulk of the total luminosity is emitted at the high-energy part of the spectrum. 
 For the pure SSC case shown in Fig.~\ref{fig2} (dashed lines), the
 SED peaks at approximately $\epsilon_{\rm ssc, \nt}= (\Gamma/(1+z))(4/3)^{\nt}b \gamma_0^{2\nt +2}\mel c^2 \simeq 2$~MeV or $\sim 5\times10^{20}$~Hz.
 {As already mentioned in Sect.~2.2, the succession of higher than $\nt$ SSC photon
 generations ceases due to the Klein-Nishina effects. 
 The highest energy of the inverse Compton scattered photons should be $\epsilon_{\rm KN} \sim \Gamma\gamma_0
 \mel c^2$ For the values numerical example of Fig.~\ref{fig2}, we find $\epsilon_{\rm KN}=0.5$~GeV or $\sim 10^{23}$~Hz.
 Notice the abrupt cutoff of the photon spectra at this frequency.}
 \item The SED shows smooth spectral breaks at the transition between successive emission components. This effect is evident, if 
 the electron distribution is mono-energetic or spans over a narrow range of energies. 
 However, for a power-law electron distribution
 with index $s > 1$, the spectral breaks are smoothened out (see next section).
 \item For even { higher compactnesses, namely ${0.2 \le \linj\lesssim 3}$}, we verified that the spectrum 
 {tends to be a power-law with }
 spectral index $\beta_{\rm cc} \sim 0.8$. 
 This is in agreement with the analytical estimate of eq.~(\ref{bcc}), {but it will become more clear with the numerical
 examples that follow (Sect.~3.2.1).}
\end{enumerate}

We have shown that there is a general agreement
between the analytical and numerical results regarding the 
electron cooling regimes, the luminosities of the various components, 
the transition of the system to the Compton catastrophe regime, and the respective spectral shape in the Compton catastrophe limit.
There are, however, several factors that might affect the spectral index $\beta_{\rm cc}$, such as the distribution
of electrons at injection (mono-energetic versus power-law) and photon-photon absorption, which redistributes the energy from the high-energy part
 of the spectrum to lower energies through an electromagnetic cascade. As the analytical treatment has its limitations, in 
 what follows, we investigate the role of these factors numerically for a pure SSC case ($\lex=0$).

\subsubsection{Effects of the electron power-law index}
So far, we have presented multi-wavelength spectra obtained for mono-energetic injection of electrons.
However, the spectral index $\beta_{\rm cc}$ may depend on
the power-law shape of the electron distribution at injection. 
To test this hypothesis, we consider 
here the case of a power-law injection of electrons (with $\gamma_{\min} \le  \gamma \le \gamma_{\max}$) 
i.e. $Q_{\rm e} \propto \gamma^{-s}H(\gamma-\gamma_{\min})H(\gamma_{\max}-\gamma)$,
with $\gamma_{\min}=1$, $\gamma_{\max}={10^{2.2}}$.
We use three indicative values for the power-law index ($s=3, 2,1$) that
correspond to electron energy spectra ranging from soft to hard.
For each case, we  {calculated the spectral for 
five  values of the injection compactness starting with $\linj = 0.01$, and increasing
$\linj$ over its previous value by $+0.5$ in logarithm. 
Other parameters used are: $B=10$~G, $\rb=10^{16}$~cm, leading to} $\lb=3\times10^{-2}$.
In order to isolate the effects that a different power-law index has on the photon spectra,
we artificially switched off the terms related to photon-photon absorption and synchrotron self-absorption 
from eqs.~(\ref{ne1}) and (\ref{ng1}).
\begin{figure}
\centering
\includegraphics[width=0.48\textwidth]{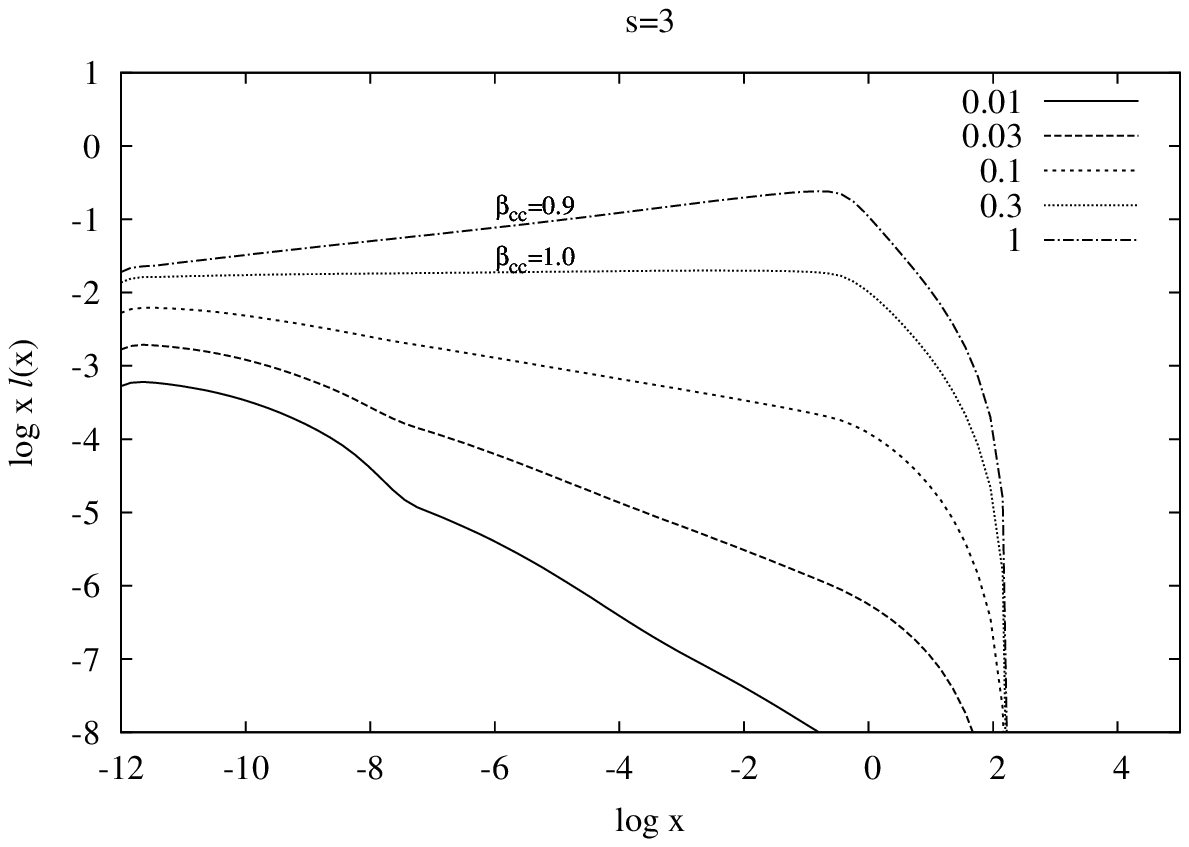}
\includegraphics[width=0.48\textwidth]{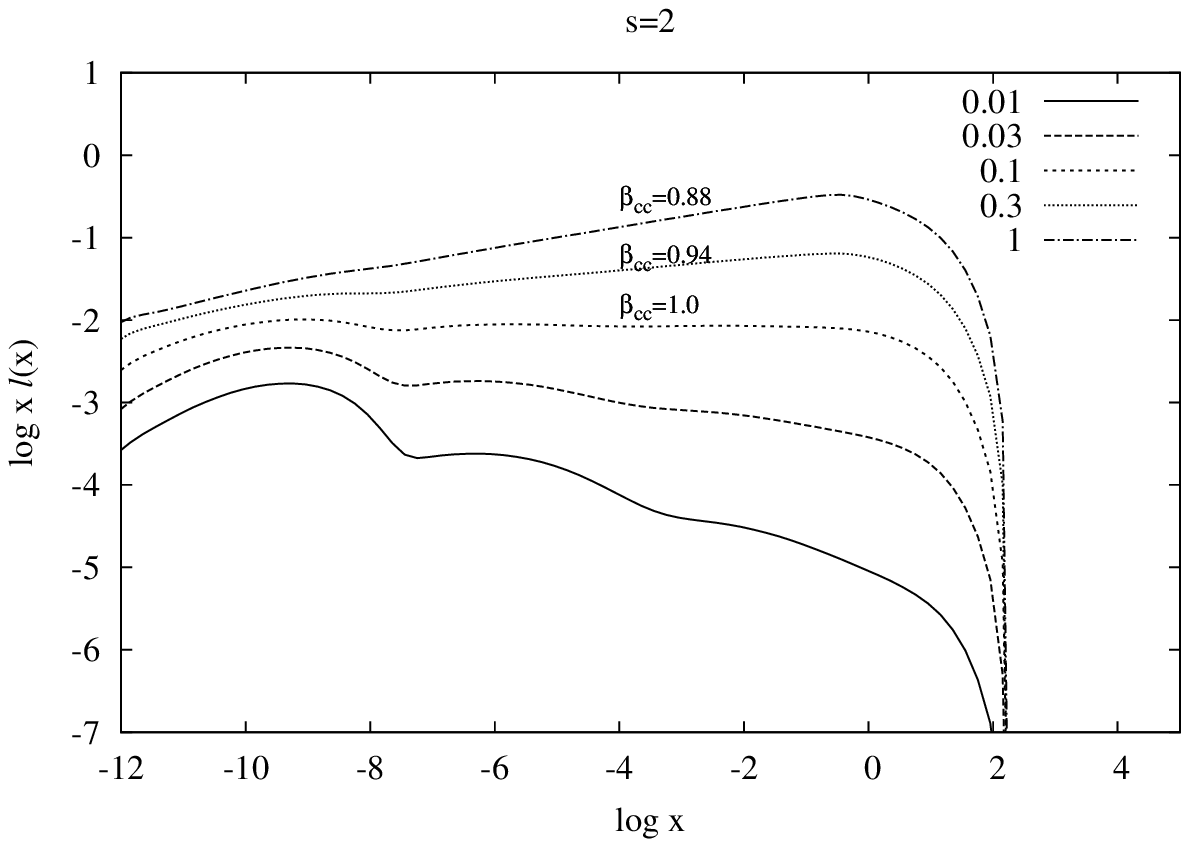} 
\includegraphics[width=0.48\textwidth]{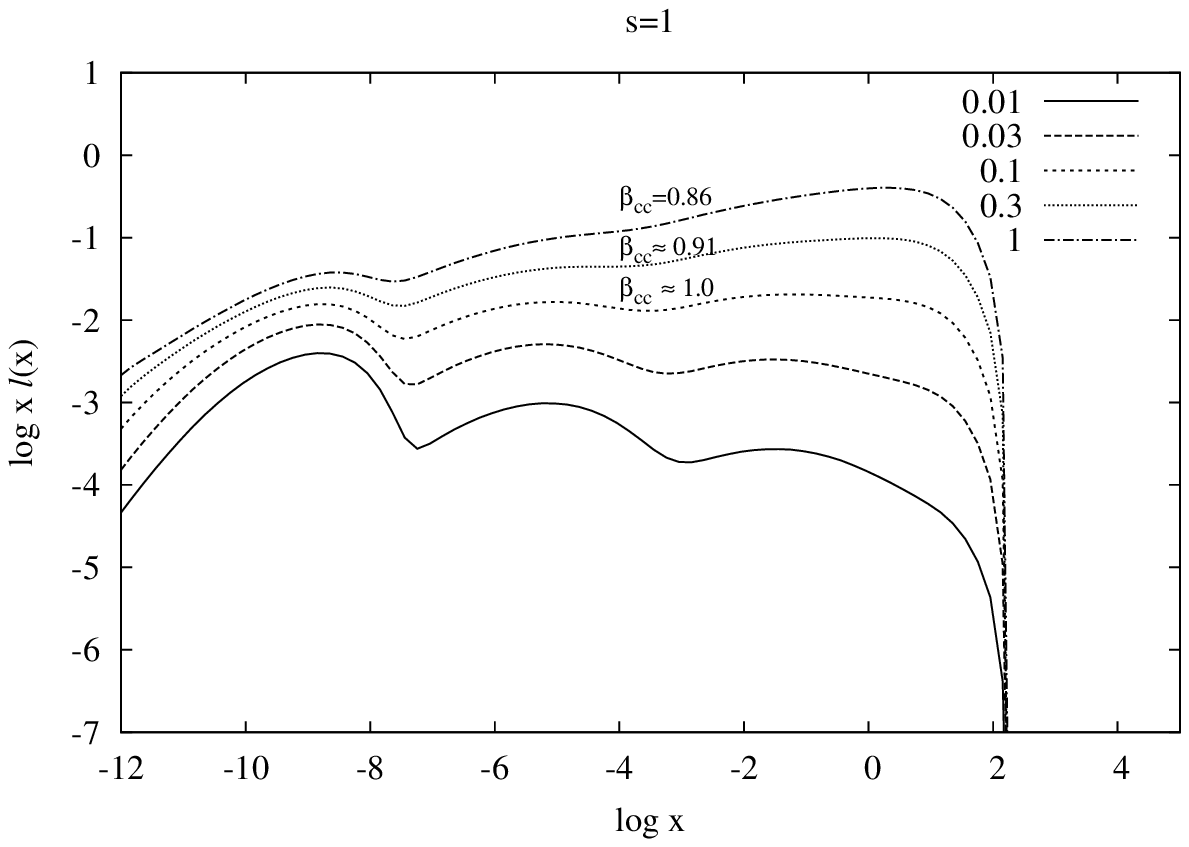} 
\caption{Steady-state multi-wavelength photon spectra obtained for a power-law injection of electrons
with index $s=3$ (top), $s=2$ (middle) and $s=1$ (bottom).  Different {types of lines} correspond
to the values of $\linj$ marked on the plot. 
Also marked is the  value of the spectral index $\beta_{\rm cc}$, whenever this can be defined. {In all panels,
x is the photon energy in $\mel c^2$ units and $\ell(x)$ is the differential photon compactness.}}
\label{fig3}
\end{figure}
\begin{figure}
\centering
\includegraphics[width=0.48\textwidth]{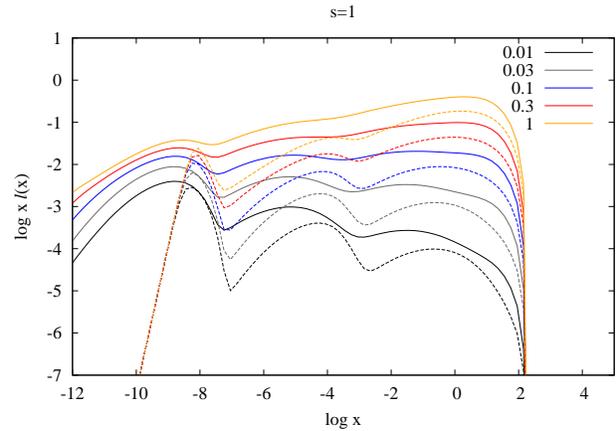} 
\caption{Same as in bottom panel of Fig.~\ref{fig3} but with the inclusion of synchrotron self-absorption as a sink term
for photons (dashed lines). 
The spectra obtained without synchrotron self-absorption are plotted with solid lines.}
\label{ssa}
\end{figure}
The results for $s=3,2$ and 1 are presented in the top, middle and bottom panels of Fig.~\ref{fig3}.
In all cases,  we show the total (synchrotron and inverse Compton) emission {in terms of photon compactness.}
Different  {types of lines} correspond to different $\linj$ marked on the plot. In all cases the steady-state electron
distribution has cooled efficiently down to $\gc\sim \gamma_{\min}\ll \gamma_0$.

For low enough electron compactnesses, {we recover the well known result \citep[e.g.][]{blumenthalgould70} that 
the spectral shape of the SED depends on the power-law index of the electron distribution} 
(see e.g. spectra for $\linj=0.01$ {in all panels} of Fig.~\ref{fig3}). 
However, as the system is pushed in the Compton catastrophe
regime ($\linj \gtrsim 0.3$), the SEDs obtain a 
{broken} power-law shape with a 
{break at $\sim \mel c^2$ as measured in the comoving frame. 
The spectral index below the break energy 
is universal, in the sense that, it is no more 
related to the power-law index
of electrons at injection, but depends
only on  $\linj, \lb$ and $\gamma_0$ {(for a discussion on the characteristic Lorentz factor
$\gamma_0$ in the case
of power-law injection, see \cite{bjornsson00}).}
However, the photon spectra for (comoving energies) $\gtrsim \mel c^2$ are steep and their 
spectral index is related to the power-law index of electrons at injection. 
The broken power-law shape and the steepening of the spectrum above the break energy
are both results of the Klein-Nishina cutoff effect (for a relevant discussion,
see \cite{zdziarskilamb86}).} 
{We note that the spectra above 
the high-energy break should appear even steeper, had the photon-photon absorption be taken into account, while 
$\beta_{\rm cc}$ would be marginally affected by the development of pair cascades (see Sect.~\ref{processes}).}

The numerically derived values of $\beta_{\rm cc}$, whenever they can be defined, are marked on the plot, and 
are in {relative} agreement with the value derived by eq.~(\ref{bcc}).
Moreover, we find that the photon spectra in the Compton catastrophe regime become harder as the electron compactness increases.
In particular, a one order of magnitude increase in $\linj$ leads to a larger $\beta_{\rm cc}$  by $\sim$0.12 units.
{The weak dependence of $\beta_{\rm cc}$ on $\linj$} is in full agreement
with the analytical prediction (see eq.~(\ref{bcc})).
On the contrary, a one order of magnitude change in $\linj$, before the source enters the Compton catastrophe regime, 
results in significant changes of the photon spectral shape (see Fig.~\ref{fig3}). 
The sensitivity of the multi-wavelength photon spectra on $\linj$, before and after the transition 
to the Compton catastrophe regime, is also reflected by the different 
density of curves in the panels of Fig.~\ref{fig3} as $\linj$ increases.

It is noteworthy that the photon spectra we obtain in the Compton catastrophe regime
are not as hard as those predicted by \cite{zdziarskilamb86}, which have a spectral index $\sim 0.3$
for even lower values of the electron compactness. After excluding a numerical error 
as the reason of this discrepancy (see Appendix~\ref{app-comp}), we argue that the hardness of the photon
spectra in the Compton catastrophe regime depends also on the nature of the seed photons used in the scatterings.
Here, we study the effects of repeated inverse Compton scatterings of the synchrotron photons by relativistic electrons, whereas
\cite{zdziarskilamb86} considered a fixed black-body photon field.

\begin{figure}
\centering
\includegraphics[width=0.49\textwidth]{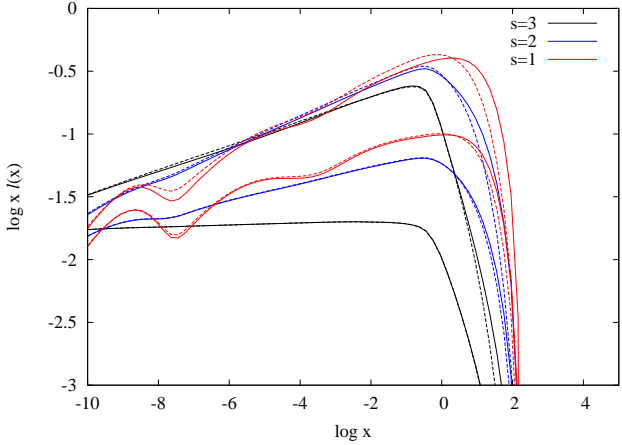} 
\caption{Same as in Fig.~\ref{fig3} but with the inclusion of photon-photon absorption (dashed lines).
The spectra obtained without photon-photon absorption are plotted with solid lines. For each value of $s$,
the lower and upper curves are obtained for $\linj=0.3$ and 1, respectively.}
\label{gg}
\end{figure}

\subsection{Effects of additional processes}
\label{processes}
We turn now to { discuss additional processes} that were not included so far and can influence the system. 
\begin{enumerate} 
{
 \item Synchrotron self-absorption: 
 for the parameter regime we are interested in, this process acts as a loss term for low-energy photons (see eq.~(\ref{ng1})) and 
reduces the number density of synchrotron photons, which are the targets for the first SSC photon
generation. It may, therefore, affect the spectral slope in the Compton catastrophe regime.
In order to test this, we calculated numerically the steady-state photon spectra for the 
same cases shown in Fig.~\ref{fig3}; the case of $s=1$ is exemplified in 
Fig.~\ref{ssa}.  In general, 
inclusion of synchrotron self-absorption makes the multiple SSC components more pronounced for low $\linj$, while
in the Compton catastrophe regime the spectrum is characterized by smaller $\beta_{\rm cc}$. 
Qualitatively speaking, this is expected, since synchrotron self-absorption removes low-energy photons
that would be, otherwise, multiply up-scattered and contribute to the high-energy emission of the source.
Quantitatively though, the effect is not
significant, since in all cases $\beta_{\rm cc}$ changed 
no more by a factor of 0.1-0.2. For example, for $s=1$ and $\linj=1$ (Fig.~\ref{ssa})
we obtain $\beta_{\rm cc}=0.8$ instead of 0.9 after the inclusion of synchrotron self-absorption. 
\item Photon-photon absorption: inclusion of this process,
makes the high energy part of the spectrum, i.e. 
at $x\gtrsim 1$, steeper. This is exemplified in Fig.~\ref{gg}, where
we show the photon spectra for $s=1,2$ and $3$ calculated with (dashed lines) and
without (solid lines) photon-photon absorption. For each value of $s$, we plot 
the spectra for $\linj=0.3$ and 1 are (from bottom to top). Only for $s=1$
and $\linj=1$ starts photon-photon absorption to have a significant impact
on the high-energy part of the spectrum. Note, however, that the 
injected secondary pairs do not modify the spectrum below the break. 
This becomes important only  for $\linj \ge 10$ \citep[see e.g. Fig.~2 in][]{svensson94}.
In short, for the parameter regime where our analysis is valid
the effects of photon-photon absorption are minimal.}
\end{enumerate}

{ Finally, other processes that might be important in optically thick
plasmas such as photon down-scattering on  electrons, electron-positron
annihilation, Coulomb collisions and bremsstrahlung turn out to
be negligible as the Thomson optical depth in all our cases is always
much less than unity.}

 \section{Relevance to astrophysical sources}
\label{astro}
We have shown that the multi-wavelength photon spectra in the Compton catastrophe regime obtain a universal power-law shape that peaks at 
{observed} energies $\gg \mel c^2$, where
most of the power is emitted. In contrast to the photon spectra obtained for low values of the electron compactness, where
one can link the observed spectral index with the power-law index of the electron distribution, in the Compton catastrophe regime
this connection  can no more be established. 
The same applies for the EC emission, which for high enough $\linj$ is hidden below the higher order SSC components.
These  properties make the photon emission in the Compton catastrophe unique.

For the derivation of the analytical relations presented in Sect.~\ref{AR}, 
we adopted the most general framework {as} possible, i.e. we did not attempt to specify the astrophysical source.
{The results regarding the observed bolometric luminosities of the various spectral components are obtained for
an emitting region } that is relativistically moving towards the observer
and is embedded in a spherical and isotropic (in the rest frame of the galaxy) 
external photon field. {These analytical relations are valid as long as the relations $\uex \simeq \Gamma^2 u_{\rm ex, \star}$ and 
$\epsilon_{\rm ex} \simeq \Gamma \epsilon_{\rm ex, \star}$ hold.}
In this section, we  apply the analytical results of Sect.~\ref{AR}
to the high-energy emitting region of either a blazar jet or a GRB jet.
\subsection{Blazar emission}
{Assuming that the blazar multi-wavelength emission (from the infrared (IR) wavelengths up to $\gamma$-ray
energies) arises from one region, then the observed spectrum cannot be reconciled with that expected
in the inverse Compton catastrophe limit. 
One example is the absence of an excessively high X-ray flux from radio-loud blazars, whose radio cores exhibit 
extreme apparent brightness temperatures. This may exclude the possibility of a source being in the  inverse Compton
catastrophe limit (e.g. \citealt{ostorero06, agudo06}). 
If, however, the one-zone assumption proves to be wrong, i.e. the IR-optical and $\gamma$-ray
emission are produced in different locations, 
then it is still possible to explain the high-energy emission in terms
of a non-thermal pair cascade \citep[e.g.][]{stern06, stern08}. 

In the following, we adopt the one-zone framework for the blazar emission, and, as an indicative example, we
focus on blazars whose $\gamma$-ray emission is dominated by the EC component.} Thus, what follows
may be of relevance for flat spectrum radio quasars (FSRQs) and low-frequency peaked 
BL~Lacs (e.g. 3C~273 and  BL~Lacertae).
In the fiducial example of Fig.~\ref{fig2} we showed that, for sufficiently high $\linj$, the emission from
the highest order SSC photon generation may dominate that of EC component. 
Thus, in the case
of EC dominated blazar emission, it is interesting to search for the conditions that suppress the emission from highest
order SSC photon generation, 
and derive information about the properties, e.g. size and luminosity, of the external radiation field.

We present the results for the fast cooling case, since it is the regime that ensures high radiative efficiency.
The luminosity ratio between the high- and low-energy components of the blazar SED
is usually referred as the Compton dominance $q$ of the source.
In cases where the EC component dominates the emission in high-energies (X-rays up to $\gamma$-rays)
the ratio is defined as $q\simeq L_{\rm EC}^{\rm obs}/L_{\rm syn}^{\rm obs}\simeq \uex/\ub$ (e.g. \citealt{sikora09})
or $q \simeq \lex/\lb$. 
The condition $L_{\rm EC}^{\rm obs} \ge L_{\rm ssc, \nt}^{\rm obs}$ results in 
\eqb
\frac{9}{4} q \ge \left(\frac{3\linj}{\lb} \right)^{\nt/(\nt+1)},
\label{eq1}
\eqe
where we used eqs.~(\ref{fast2-Lssc}) and (\ref{fast2-Lec}). 
Using the constraint of eq.~(\ref{conf6}) we find
that 
\eqb
\left(\frac{3\linj}{\lb}\right)^{\nt/(\nt+1)} > \frac{3}{4\lb\gamma_0},
\label{eq2}
\eqe
which when combined with  the relation (\ref{eq1}) results in $\rex < r_{\max}$ with $r_{\max}$ given by 
\eqb
\label{eq3}
r_{\max} = 3.2 \times 10^{17} \ {\rm cm} \ \left(\Gamma_1^2 \eta_{-1} L_{\rm d, 46} \right)^{1/2} \left( r_{\rm b, 16} \gamma_{0, 2} \right)^{1/2},
\eqe
where $\Gamma$ is the Lorentz factor of the emission region. 
In the above, we introduced the notation $F_{x}=F/10^x$ in csg units, unless stated otherwise. 
We also assumed that the source of the external radiation is the reflected blazar's disk luminosity  ($L_{\rm d}$)
at the broad line region (BLR). The latter, for the purposes of this study,
was taken to be spherical with radius $\rex$ (see also Fig.~\ref{sketch}).
In the derivation of (\ref{eq3}) we, thus, 
assumed $L_{\rm ex}=\eta L_{\rm d}$ and used $\eta=0.1 \eta_{-1}$ and $L_{\rm d}=L_{\rm d, 46}/10^{46}$~erg/s
as typical values (see e.g. \citealt{boettcherreimer13, nalewajko14}).

Relation (\ref{eq3}) shows that the inverse Compton catastrophe
may be avoided, if the external source of radiation, the BLR in our case, 
has a sub-pc scale size ($\rex \sim 0.1$~pc)\footnote{Interestingly, the BLR radius is typically found to be 
$\sim 0.1$~pc (see \citealt{nalewajko14}, and references therein).}.
For a given $L_{\rm ex}$, a pc-scale BLR on the other hand, is dilute enough as to favour the SSC emission and even lead the source
to the Compton catastrophe limit for sufficiently high $\linj$.
In general, the upper limit we derived for  $\rex$ becomes {even} more stringent for lower reflection efficiencies $\eta$
and less luminous disks. {Although} a larger emission region and/or the injection of electrons with higher  Lorentz factors, push
the upper limit close to the pc-scale, the dependence of $\rex$ on both parameters is relatively weak ($r_{\max} \propto (\gamma_0 \rb)^{1/2}$).

The aforementioned results suggest  that the inverse Compton catastrophe  does not typically occur in these sources.
Before closing this paragraph we note that higher order SSC photon generations ($\nt\ge 2$) may
become relevant for $\gamma$-ray emitting radio galaxies, such as Cen~A and M~87, even if the sources
are not in the extreme regime of the Compton catastrophe. In the particular case of Cen~A, \cite{petrolefa14}
studied in detail the role of the second SSC photon generation on the high-energy emission detected from the core of Cen~A.

\subsection{GRB emission}
{ Amongst the various proposed mechanisms for the GeV emission of GRBs\footnote{{Although the GRB GeV emission
became a topic of interest after the Fermi-LAT detections \citep{ackermann13}, 
it has been already observed with CGRO/EGRET \citep{hurley94, gonzalez03}.}}
}, the inverse Compton scattering has been discussed in several studies 
\citep[e.g.][]{Dermer+00,GuettaGranot03,GranotGuetta03,PeerWaxman04,Beloborodov05,fanpiran06,GuptaZhang07, andoetal08, fanpiran08}. 
External inverse Compton  from various photon sources has been also 
considered by many authors \citep[e.g.][]{Shaviv95, ShavivDar95,giannios08} beginning
with \cite{Shemi94}, while some of its implications when applied as an up-scattering process of the hot
cocoon radiation have been addressed by \cite{toma09} and \cite{kumarsmoot14}.

{ Regarding the  (sub)MeV GRB emission, the volume of literature
is huge \citep[e.g.][]{piran04, peer06, beloborodov10, vurm11, giannios12}, with each
study reproducing at a certain degree the basic observed properties.
Here, we test the alternative hypothesis
of an external Compton origin, since it offers a physical setup for illustrating
the principles of the inverse Compton catastrophe and the generality of the analytical
framework we developed in Sect.~\ref{AR}. In what follows,
we will show how the inverse Compton catastrophe alone, can
be used to reject this hypothesis.
}

We begin with a short description of the GRB jet model that we will use
to accordingly define quantities, such  $\lb$ and $\rb$, that appear
in the analytical expressions
of Sect.~\ref{AR}. 
Let us consider a GRB flow of kinetic (isotropic equivalent) luminosity
$L_{\rm k}$ and bulk Lorentz factor $\Gamma$. 
When the jet reaches a distance $\rem$ a substantial fraction
of its luminosity is dissipated internally leading to acceleration of electrons.
A fraction of  the dissipated energy results in the prompt
GRB emission. We assume that the GRB emission takes place
in the optically thin to Thomson scatterings
part of the GRB jet (e.g. \citealt{piran99, meszarosrees00, giannios12}),  namely 
\eqb
\rem \gtrsim r_{\rm T} \equiv 10^{14} \ {\rm cm} \frac{L_{\rm k,53}}{\Gamma_2^3},
\label{eq4}
\eqe
where $\rem$ is measured in the galaxy rest frame. The size of the emission region, as measured in the respective rest frame, 
is then parameterized as $\rb \simeq \rem/\Gamma \gtrsim 10^{12} L_{\rm k, 53} \Gamma_2^{-4}$~cm.
An estimate of the magnetic field in the emission region is given by
\eqb
B= \left(\frac{\epsilon_{\rm B} L_{\rm k}}{c} \right)^{1/2} \frac{1}{ \rem \Gamma}.
\label{eq5}
\eqe
where $\epsilon_{\rm B}$ denotes the ratio of the Poynting luminosity to the 
jet kinetic luminosity. The magnetic compactness of the emission region
is written as
\eqb
\lb = \frac{\sth \epsilon_{\rm B}L_{\rm k}}{8\pi \mel c^3 \rem \Gamma^3}.
\label{eq6}
\eqe
The external radiation field, which will be up-scattered by the electrons
in the emission region of the GRB jet, is left undefined,
apart from the fact that is assumed to be spherically symmetric and isotropic in the rest frame of the explosion.
We also assume that the emission region is embedded in it:
\eqb
\rem = 10^{14} \ {\rm cm} \frac{L_{\rm k, 53}} {\Gamma_2^3} \lesssim \rex.
\label{rex}
\eqe
This can be considered as  
the most efficient case of a more general scenario where  $\rem > \rex$, since the energy
density of external photons is higher in the comoving frame by a factor of $\Gamma^2$.
In the limit where $\rem \gg \rex$, the energy density of external photons will appear in the comoving frame of
the ejecta de-boosted \citep{dermerschlickeiser93}.

Two are the observational constraints that we will use in our analysis:
\begin{enumerate}
\item The peak energy of the GRB spectrum $\epeak \sim 0.2$~MeV \citep{gruber14},{which can be used to }
express the energy of the external photons as
  \eqb
 \eex \approx 10 \ {\rm eV}\left(\frac{\epeak}{0.2 \ {\rm MeV}}\right) \left(\frac{1+z}{2}\right) \frac{1}{\Gamma_2^2 \gamma_0^2}.
 \label{epeak}
 \eqe
 Because the energy of external photons appears boosted in the comoving frame by a factor of $\Gamma$, the above
 relation sets a strong constraint on the product $\gamma_0^2 \eex$, {at least in the regime of $\gamma_0 \gtrsim 3$
 where our analysis
 is applicable.} 
 \item The GRB energy flux in the energy range 10~keV-1~MeV is $F_{\gamma}^{\rm obs} \sim 3 \times 10^{-7}$~erg cm$^{-2}$~s$^{-1}$ \citep{gruber14}. \\
 In the scenario of an EC origin of the prompt GRB emission, we thus require $L_{\rm EC}^{\rm obs} = 4\pi D_{\rm L}^2 
 F_{\gamma}^{\rm obs} \simeq 1.5\times 10^{51}$~erg/s, for $D_{\rm L}=6.7$~Gpc ($z=1$). In what follows,
 we normalize $L_{\rm EC}^{\rm obs}$ with respect to $10^{51}$~erg/s.
\end{enumerate}
{These observational constraints in addition to the principles of the inverse 
Compton catastrophe can be used to exclude an EC origin of the (sub)MeV GRB emission. 
We demonstrate this in detail for the slow cooling regime even though one expects the emitting region 
to be fast cooling.  The expressions are less complicated in this regime and 
similar conclusions can be reached for the fast cooling regime, although a lengthier derivation
is required\footnote{{In the fast cooling regime, the inverse Compton origin of the $\sim$MeV GRB emission
is also challenged  on a more general ground: the up-scattering of a black-body seed photon field  
by cooled electrons ($n_{\rm e} \propto \gamma^{-2}$) results in much softer $\gamma$-ray spectra than typically observed (e.g. \citealt{preece98, preece00}).}}.
}
Solving eq.~(\ref{epeak}) with respect to $\gamma_0$ and substitution into 
eq.~(\ref{lum-ec-slow}) results in
\eqb
L_{\rm ex} \linj > 10^{43}~{\rm erg/s}\ \frac{L_{\rm EC, 51}^{\rm obs}}{\Gamma_2^3}
\left(\frac{0.2 {\rm MeV}}{\epeak}\right)^{1/2}\left(\frac{2}{1+z}\right)^{1/2}\left(\frac{\eex}{10 {\rm eV}} \right)^{1/2},
\label{Lex}
\eqe
where we made also use of eq.~(\ref{rex}). 

Relations (\ref{rex}), (\ref{epeak}) and (\ref{Lex}) describe an external source of photons with the following properties:
$\rex > 1500 R_{\odot}$, $\eex \sim 1$~eV for $\gamma_0=3$ and 
$L_{\rm ex} > 10^9 L_{\odot}/\linj$, where $R_{\odot}=7\times 10^{10}$~cm and $L_{\odot}\simeq 4\times 10^{33}$~erg/s
stand for the solar radius and luminosity, respectively. We discuss each of the requirements below.

The typical seed photon energy lies in the optical energy range for mildly relativistic electrons, while the injection of 
relativistic electrons with $\gamma > 10$ would push the typical energy below the far-IR regime. 
Such low energy photons are not, however, common in GRB environments.  For example,
in the collapsar model for GRBs \citep{woosley93}, the burst is related to the
core-collapse of a massive star, typically that of a Wolf-Rayet star  (e.g. GRB 980425 \cite{galama98}; GRB 030329 \cite{hjorth03}),
whose emission usually peaks in the far-UV/soft X-ray regime. 
A companion star would typically be of similar size and hence its emission is also expected to be dominated by this energy band.

The fact that $\rex \gg R_{\odot}$ suggests that the seed photons for EC scattering
cannot be directly provided by the progenitor star.
A spatially extended source is, therefore, more plausible. 
One possibility is that photons from the collapsar are reflected at the strong wind
of the massive star. In this case, the energy density of the reflected photons would still appear boosted
by a factor of $\Gamma^2$ in the comoving frame of the emitting region. However, for  
typical values for the mass loss rate and the wind velocity  of the progenitor \citep{chevalier00},
the Thomson optical depth for scattering is given by
\eqb
\tau_{\rm T}=\frac{\sth \dot{M}_{\rm w}}{4\pi \rex \mpr V_{\rm w}}
\simeq 2\times 10^{-3}\frac{\dot{M}_{\rm w}}{10^{-5} M_{\odot} / {\rm yr}}\frac{10^{8} {\rm cm/s}}{V_{\rm w}}\frac{10^{14} {\rm cm}}{\rex},
\eqe
which makes this scenario less efficient. Even if there were parameters that led to $\tau_{\rm T}\sim 1$, the required luminosity of the external photon field
would still be many orders of magnitude larger than typical luminosities of massive stars. 
The cocoon's emission is stronger and even with such a small optical depth 
enough energy might be reflected back to serve as seed for external inverse Compton. 
However, the typical  energy of the cocoon's photons is $\sim 1$~keV, which is much too high to serve as a seed for scattering to the sub-MeV $\gamma$-rays.
 
The luminosity constraint is relaxed, if we assume that \linebreak $\linj \gg 1$. 
In principle, this is a viable assumption, but
pushes the source into the Compton catastrophe limit.
Roughly speaking, we showed that for $\linj \gg 1/4\gamma_0$, the luminosity of the $(i+1)$-th SSC generation becomes
larger than preceding one (see Figs.~\ref{par-space} and \ref{beta}), and eventually leads to 
$L_{\rm ssc, \nt}^{\rm obs} \gg L_{\rm EC}^{\rm obs}$. 
Since, the peak energies of the two components
differ, the final $\gamma$-ray emission will
be radically different from the Band-like GRB observed spectrum \citep{band93}.
More specifically, the electron compactness {that leads} to $L_{\rm ssc, \nt}^{\rm obs} > L_{\rm EC}^{\rm obs}$ should be higher than 
\eqb
\linj > \frac{1}{4\gamma_0}\left(\frac{\rem}{\rex}\right)^{2/\nt} \left(\frac{2\Gamma^4 L_{\rm ex}}{\epsilon_{\rm B} L_{\rm k}}\right)^{1/\nt},
\label{eq8}
\eqe
where we used eqs.~(\ref{lum-ssc-slow}), (\ref{lum-ec-slow}) and (\ref{eq5}). The above relation may also be written
as 

\begin{eqnarray}
\linj &>& 2^{\frac{1-2\nt}{1+\nt}}\Gamma_2\left( \frac{\eex}{10 {\rm eV}}\right)^{1/2}
\left(\frac{0.2 {\rm MeV}}{\epeak} \right)^{1/2} \left(\frac{2}{1+z}\right)^{1/2}
\left(\frac{\rem}{\rex}\right)^{\frac{2}{\nt+1}} \nonumber \\
&&\left(\frac{L_{\rm EC, 51}^{\rm obs}}{\epsilon_{\rm B, -1} L_{\rm k, 52}}\right)^{\frac{1}{\nt+1}},
\label{eq9}
\end{eqnarray}
where we made use of  eq.~(\ref{epeak}) and relation (\ref{Lex}).
Even if $\nt=1$, the numerical factor at the right hand side of the  relation (\ref{eq9}) is of order unity.
Thus, if we try to relax the luminosity constrain by using $\linj \gg 1$, we find that the source is driven into the 
Compton catastrophe regime and at the same time $L_{\rm ssc, \nt}^{\rm obs} \gg L_{\rm EC}^{\rm obs}$.

\begin{figure}
\centering
\includegraphics[width=0.48\textwidth]{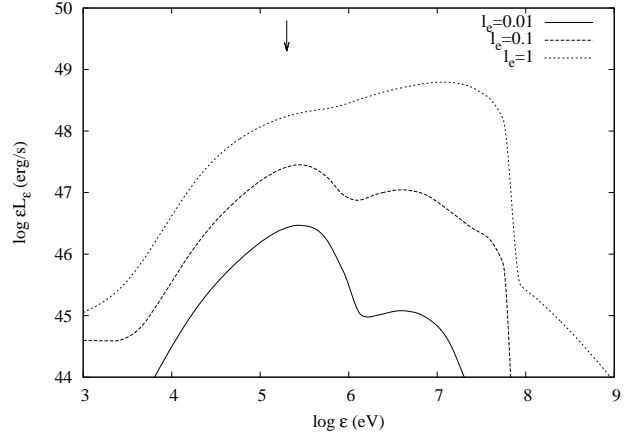}
\caption{Photon spectra in 1~keV-1~GeV energy range obtained in the EC scenario for the prompt GRB emission. Different
curves correspond to different values of the electron compactness (see legend).
{The arrow denotes the position of the observed peak energy, which 
is taken to be $\epsilon_{\rm p, obs}=0.2$~MeV.} For the rest of the parameters,
see text.}
\label{example}
\end{figure}

We demonstrate the onset of the inverse Compton catastrophe through an indicative example, 
where we used the following parameters: $z=1$, 
$L_{\rm k}=10^{53}$~erg/s, $\eb=10^{-5}$, $\Gamma=100$, $\rem=10^{14}$~cm, $\rb\simeq \rem/\Gamma=10^{12}$~cm,  and
$\gamma_0=3$. We considered a fiducial photon field with $\eex=10$~eV, $\rex\sim \rem$, and $L_{\rm ex}=3\times 10^{44}$~erg/s. 
The magnetic and external photon compactnesses for these values are $\lb=0.01$ and $\lex=0.07$. For the electron
compactness, we adopted {three} values ranging from $10^{-2}$ up to {1}; given the other parameters, the maximum value of $\linj$ 
results in $L_{\rm EC}\sim 10^{49}$~erg/s (see eq.~(\ref{lum-ec-slow})).
The high-energy spectra, which correspond to one GRB pulse with duration $\dt\sim \rb/c\Gamma \simeq 0.3$~s, are shown 
in Fig.~\ref{example}. For $\linj \le 0.1$, the EC component dominates the $\gamma$-ray emission, at the cost, however, of
low luminosity. Increase of $\linj$ transfers the $\gamma$-ray power from the EC to the highest SSC component. Thus, although
the power is $\sim 10^{51}$erg/s for $\linj=100$, it is emitted at $\gg $~MeV energies. 
Notice that we have already used a high fiducial value for $L_{\rm ex}$.

Concluding, we have shown, using   simple arguments, that there are no reasonable physical parameters that lead to $L_{\rm EC}^{\rm obs} \sim 10^{51}$~erg/s
and at the same time suppress the Compton catastrophe effect.
\section{Discussion}
\label{discussion}
The {\sl inverse Compton catastrophe} is essentially a runaway 
that transfers radiative power from low energy to high energy photons. 
It is applicable to leptonic plasmas where relativistic electrons radiate seed photons
by synchrotron and/or inverse Compton scattering, followed by further inverse Compton scatterings. 
If the source is compact enough in relativistic electrons,
one can find parameter regimes, 
where the energy density of synchrotron photons becomes
larger than the magnetic one.
This sets a threshold condition above which 
each successive generation of radiated photons due to the inverse Compton process
becomes more important to electron losses than its previous one. This is a non-linear process, since
it depends on internally produced photons, and could lead into
an exponentiation of electron losses, hence the
term `inverse Compton catastrophe'.

The aim of this  paper is to examine the physical requirements for an {optically thin to Thomson scatterings}
source to
enter the inverse Compton catastrophe regime and to calculate  the expected radiative signatures in such a case.
We have considered a one-zone model, where relativistic electrons are injected at a constant rate into a magnetized spherical 
source  that is immersed inside an external photon bath. We calculated, using
standard expressions for the energy losses and radiative processes {in the relativistic regime}, the evolution
of the system until it reaches a steady state. 
Because of the intrinsic non-linearity of problem, we restrict our analytical 
investigation to asymptotic regimes. The results of this analysis are, however, 
enlightening, and in rough agreement with our numerical ones, which are based on 
the simultaneous solution of the coupled kinetic equations for electrons and photons. 
This allows us to treat in detail
the non-linear effects, which become central during the transition to the inverse 
Compton catastrophe regime.

There are various source parameters that play an important role for the onset of inverse Compton catastrophe.
As can be deduced from the analytical treatment presented in Sect.~2, the key role  belongs to 
the compactness of the injected electrons. An inspection of eq.~(\ref{utot}) reveals that as long as this is 
smaller than the magnetic compactness or the external photon one, then the system is linear and its solution 
leads to the standard expressions for the  electron and photon distributions  in the `fast' or `slow' cooling regimes.
On the other hand, if the electron compactness increase, an increasing part of the losses is 
shifted to the internally produced synchrotron and inverse Compton photons. We have set the 
threshold for the inverse Compton catastrophe
to be that value of the electron compactness that makes the electron losses due to external agents (magnetic and external photon fields)
to be equal to the internal ones (synchrotron and Compton). Above that value the system becomes non-linear.

A more sublime parameter for the onset of the inverse Compton catastrophe 
is the characteristic injection energy  of the electrons ($\gamma_0$), under the assumption that these are mono-energetic.
This is related to the fact that the inverse Compton catastrophe enters its full {\sl modus operandi} only when in the
Thomson regime.  The exponential 
photon growth in the source is naturally suppressed by the Klein-Nishina effects.
Thus, as  eq.~(\ref{NT}) reveals, only small values of $\gamma_0$ can produce several Compton 
generations before the Klein-Nishina cutoff is reached. 


As we have shown in Sect.~2, one can find parameters, {albeit limited}, that make the system enter the inverse Compton
catastrophe limit, while the electrons do not cool efficiently (slow cooling regime). 
This is in contrast to the standard notion, according to which, the inverse Compton
catastrophe is alway associated with efficient electron cooling.
So, at least according to our definition, the inverse Compton catastrophe
is a misnomer because it does not necessarily 
imply  catastrophic electron losses. 
    
An interesting feature of a source that is deep in the non-linear
regime, i.e. when the electron injection compactness dominates over all the
other compactnesses, is its multi-wavelength photon emission. In Sect.~3 we have shown that
the photon spectra obtain an almost universal broken power-law shape, and it extend 
up to the maximum electron energy -- see Fig.~\ref{fig3}.  The break  occurs at $\sim \mel c^2$ in the
rest frame of the source and is related to the Klein-Nishina suppression. 
The spectral index below the break energy does not depend on the power-law index of the injected electron
distribution, as is the case in the linear regime (both in the slow and fast cooling limits). 
In Sect.~2 and 3 we have shown that the spectral index in the Compton catastrophe regime depends on 
the electron and magnetic compactness, but only logarithmically. 
Strictly speaking, the photon spectra in the Compton catastrophe regime
do not saturate. Yet,  they are largely independent of the specifics of the external photons, while
their dependence on the electron and magnetic compactnesses is very weak. Thus, 
one can draw some analogies to the thermal Comptonization process, where the photon spectrum saturates into a
specific power-law up to the maximum possible energy.  

The multi-wavelength spectra from astrophysical sources, such as blazars and GRBs, do not
generally resemble these expected in the Compton catastrophe regime, 
which as we have shown 
bear some unique features. {Assuming that the multi-wavelength emission of 
compact non-thermal sources originates
from one region one can then apply the principles of the inverse Compton catastrophe
to constrain their parameter space}.
The need of large luminosities when combined with the inferred small radii
push the compactnesses of these sources 
to high values, and in order for the Compton catastrophe to be avoided, special conditions are required. 
We present two indicative applications in Sect.~\ref{astro}.  For the case of blazars we set
an upper limit to the size of the BLR region. In the
GRB case, we have shown how one can use the Compton catastrophe to make a rather robust argument about
ruling out an external Compton origin of the prompt 
(sub) MeV emission.

We have studied the radiative signatures of compact sources in the inverse Compton
catastrophe regime in a time-independent way by focusing on the steady-state
photon spectra. These obtain a power-law shape that is the combined
result of multiple inverse Compton scattered photon generations. As the flux
of each underlying photon generation depends on the various parameters in a different way,
time variations in one or more parameters, such as the luminosity of injected electrons,
will have a different impact on their respective emission. The end result of the escaping
photon spectrum in such cases is non-trivial and requires 
a time-dependent approach that will be the subject of a future study.

\section*{Acknowledgments}
We would like to thank Dr. S. Dimitrakoudis for contributing the illustration in Figure 2
and Dr. Rodolfo Barniol Duran for comments on the manuscript. 
{We would also like to thank the anonymous referee for comments and suggestions that helped clarify the manuscript.}
Support for this work was provided by NASA through Einstein Postdoctoral 
Fellowship grant number PF3~140113 awarded by the Chandra X-ray 
Center, which is operated by the Smithsonian Astrophysical Observatory
for NASA under contract NAS8-03060 (MP) and by an ERC advanced 
grant ``GRBs",  by the I-CORE  Program of the Planning and 
Budgeting Committee and The Israel Science Foundation grant No 1829/12 and by 
ISA grant 3-10417  (TP).

\appendix
\section[]{Energy density of higher order SSC photon generations}
\label{app1}
We calculate the energy density of the $i$-th SSC photon generation ($u_{\rm ssc,i}$) under the assumption of 
inverse Compton scattering in the Thomson regime.
A rough estimate of the number ($\nt$) of the SSC photon generations that will be produced in the Thomson regime can be found by
\eqb
y_{\nt} < \frac{3}{4},
\eqe
 where
$y_{\rm i}$ is defined as 
\eqb
y_{\rm i} = \frac{\gamma_0 \epsilon_{\rm i-1}}{\mel c^2} = b \left(\frac{4}{3} \right)^{i-1}  \gamma_0^{2i+1},
\eqe
where we inserted $\epsilon_{\rm i}/ \mel c^2 = (4/3)^{i}b\gamma_0^{2(i+1)}$.
Thus,
\eqb
\nt = \left[\frac{\log \left(\frac{1}{b\gamma_0} \right)}{\log\left(\frac{4\gamma_0^2}{3} \right)}\right],
\eqe
where the brackets denote the integer value of the enclosed expression. Assuming $B=1$~G and $\gamma_0=10$ we find
$\nt=5$, whereas for more energetic electrons, e.g.  $\gamma_0=10^3$, this reduces to $\nt=1$.

Under the $\delta$-function approximation for the synchrotron emissivity (see e.g. \citealt{mastkirk95}),
the number density of synchrotron photons is given by
\eqb
n_{\rm s}(x) = \frac{2}{3} \rb \sth \frac{\ub}{\mel  c^2} b^{-3/2} x^{-1/2} \nel\left(\sqrt{\frac{x}{b}}\right),
\eqe
where $x$ is the photon energy in $\mel c^2$ units. Using the same approximation,  the inverse Compton emissivity of a single electron
is written as 
\eqb
j_{\rm ssc,i}(x_{\rm ssc,i}) = \frac{4}{3}\sth c u_{\rm ssc, i-1}\gamma^2 \delta\left(x_{\rm ssc, i} -\frac{4}{3} \gamma^2 x_{\rm ssc, i-1}\right),
\eqe
and energy density of the $i$-th SSC photon generation is then given by
\eqb
u_{\rm ssc, i} = \frac{4}{3}\sth
\rb \int dx_{\rm ssc, i} \int d\gamma \gamma^2  \nel(\gamma) u_{\rm ssc, i-1}\delta\left(x_{\rm ssc, i} -\frac{4}{3} \gamma^2 x_{\rm ssc, i-1}\right)
\eqe
The stationary electron distribution $\nel$ is written 
as $\nel(\gamma)=\ks \delta(\gamma-\gamma_0)$ and  $\kf\gamma^{-2}$ for $\gamma \le \gamma_0$, 
in the limiting cases of {\sl slow} and {\sl fast} cooling, respectively.
The energy density of synchrotron photons and successive SSC generations
is then given by
\eqb
\label{us}
\us & = & \frac{4}{3}\sth \rb \ub G_{\rm e}, \\
u_{\rm ssc, 1} & = & \ub  \left(\frac{4}{3}\sth \rb G_{\rm e}\right)^2  \\
u_{\rm ssc, 2} & = & \ub  \left(\frac{4}{3}\sth \rb G_{\rm e}\right)^3  \\
& \vdots & \nonumber \\
u_{\rm ssc, i} & = & \ub  \left(\frac{4}{3}\sth \rb G_{\rm e}\right)^{i+1},
\label{ussc}
\eqe
where 
\eqb
G_{\rm e} = \left\{ \begin{array}{ll}
             \ks\left(\linj\right)\gamma_0^2, & {\rm slow \ cooling} \\
             \kf\left(\linj,\lex,\lb\right) \gamma_0, & {\rm fast \ cooling}
                  \end{array}
	    \right.
	    \label{ge}
\eqe
Equations (\ref{us})-(\ref{ussc}) are similar in both cooling regimes. All the information about
the evolution of the electron distribution to a steady-state under the influence of escape, injection 
and cooling enters only through $\ks$ and $\kf$, which are themselves 
functions of the basic compactnesses that describe the physical system.

\section[]{Inclusion of the EC energy density in electron cooling}
\label{uEC}
The average energy of once inverse Compton scattered external photons, as measured in the rest frame of the emission region,
is $\epsilon_{\rm EC} = (4/3) \gamma_0^2 (\Gamma \eex)$, where $\gamma_0$ is the electron Lorentz factor,
$\Gamma$ is the bulk Lorentz factor of the emission region, and $\eex$ is the characteristic energy of the external photons
as measured in the rest frame of the galaxy. From this point on, we use unprimed and starred quantities, when these are 
measured in the rest frame of the emission region and the galaxy, respectively.
In the above, we assumed that boosting of the external photon energy by a factor of $\sim\Gamma$ because of the relativistic motion of the emission region
with respect to the external source of photons.
The inverse Compton scattering of the EC photons by electrons with Lorentz factor $\gamma_0=100$ takes
place in the Klein-Nishina regime, since 
\eqb
\gamma_0 \epsilon_{\rm EC} \sim 10^7 \ {\rm eV} \left(\frac{\gamma_0}{100}\right)^3 \frac{\Gamma}{10} \frac{\eex}{1{\rm eV}} \gg \mel c^2, 
\eqe
and one can safely neglect it from the electron cooling  term (see eq.~(\ref{utot})).
However, for lower Lorentz factors, e.g. $\gamma_0=10$, 
the above relation becomes $\gamma_0 \epsilon_{\rm EC} < \mel c^2$; the first inverse Compton scattering of EC photons would 
occur in the Thomson regime, while the second scattering would marginally fall in the Klein-Nishina regime. We can therefore include 
only the first generation of EC photons in the electron cooling term.
In what follows, we will perform a similar  analysis  to the one presented in Sects.~2.2.1-2.2.2, but with
the addition of the energy density of EC photons ($u_{\rm EC}$) to the total one.

Under the $\delta$-function approximation for the inverse Compton emissivity (see Appendix~\ref{app1}),
the energy density of the EC photons is given by
\eqb
u_{\rm EC} = \frac{4}{3}\sth\rb G_{\rm e} \uex,
\label{uec}
\eqe
where $G_{\rm e}$ is defined in eq.~(\ref{ge}). 
The similiraty to the energy density of synchrotron photons
is evident, with the external photon source  playing, in this case, the role of the magnetic field (eq.~\ref{us}).
Finally, one can define the respective  photon compactness as
\eqb
\ell_{\rm EC} = \frac{4}{3}\sth \rb G_{\rm e} \lex.
\label{lec}
\eqe
\subsection{Slow cooling regime}
The steady-state electron distribution is not determined by
energy loss processes in this regime. Thus, the inclusion of $u_{\rm EC}$ in the electron cooling term should not alter the 
solution for $\nel$, which is written as $\nel(\gamma)=\ks \delta(\gamma-\gamma_0)$ with $\ks$ given by eq.~(\ref{ke-slow}). 

The total compactness of the source, in this case, is written as:
\eqb
\ell_{\rm tot}=\lb+\lex+\lb(4\linj\gamma_0) + \lex(4\linj\gamma_0) + \lb \sum_{\rm i=1}^{\nt-1} \left(4\linj\gamma_0\right)^{i+1},
\label{ltot_slow_app}
\eqe
where we made use of eqs.~(\ref{lsyn-slow}), (\ref{lssc-slow}) and (\ref{lec}). Similar to what we have shown in Sect.~2.2.1,
we may approximate the total compactness of the source as 
\eqb
\ell_{\rm tot} \approx \left\{\begin{array}{ll} 
\lb + \lex,  & \linj < 1/4\gamma_0 \\
\lex \left(4\linj \gamma_0 \right) +  \lb \left(4\linj \gamma_0\right)^{\nt},  & \linj > 1/4\gamma_0,
 \end{array}
 \right.
 \label{ltot_aprox1}
 \eqe
where for the latter, we assumed that $\lex > \lb \left(4\linj \gamma_0\right)^{\nt-2}$. 
For $\linj > 1/4\gamma_0$, the total compactness of the source is larger than the respective one in eq.~(\ref{ltot_slow}), as expected.
The slow cooling condition, i.e. $\xi >1$ or $3/ (4\linj \gamma_0) > 1$,  is satisfied if  
\eqb
\label{cons1-app}
\linj & < & \frac{1}{4\gamma_0}  \ {\rm and} \\
f_{\rm ex} & < & \frac{3}{4\lb \gamma_0},
\label{cons2-app}
\eqe 
or 
\eqb
\label{cons3-app}
\linj & > & \frac{1}{4\gamma_0}  \ {\rm and} \\
f_{\rm ex} & < & 1+ \frac{1}{4\linj \gamma_0}\left(\frac{3}{4\lb \gamma_0}-\left(4\linj \gamma_0 \right)^{\nt-1}\right),
\label{cons4-app}
\eqe 
where $f_{\rm ex}$ is defined in eq.~(\ref{fex}). 
We note that the above constraints are identical to these of Sect.~2.2.1 apart from the last one (see eqs.~(\ref{cons4}) and (\ref{cons4-app}). Inclusion
of the $u_{\rm EC}$ makes the upper limit on $f_{\rm ex}$ more stringent than before. Thus, for the same $\linj$ and $\lb$,
the external photon source should be less compact, i.e. the ratio $\lex/\lb$ should be smaller, in order to suppress electron cooling.

\subsection{Fast cooling regime}
In the fast cooling regime, the steady-state equation of electrons may be written as
\eqb
\kf \left(f_{\rm ex}\left(1+ \alpha \kf \right) + \sum_{i=1}^{\nt-1} \left(\alpha \kf\right)^{i+1} \right) =  
\frac{9\linj/\lb}{4\sth \rb \gamma_0},
\label{general-app}
\eqe
where $\alpha$ is defined in eq.~(\ref{aa}). In the above, we have also inserted 
the compactness term related to the EC photons, i.e. $l_{\rm EC} = \lex \alpha \kf$.
Same as in Sect.~2.2.2, eq.~(\ref{general-app}) can be studied in two regimes according to the value of $\alpha  \kf$:
\begin{itemize}
 \item for $\alpha \kf < 1$, the main energy loss channels for electrons are synchrotron radiation and 
inverse Compton scattering on external photons, synchrotron and EC photons. The total compactness 
can be approximated as 
\eqb
\ell_{\rm tot} \approx \lb + \lex + \lsyn +\ell_{\rm EC}= \lb f_{\rm ex}(1+ \alpha \kf).
\label{ltot_aprox2}
\eqe
The steady-state equation (\ref{general-app}) then simplifies into a second order polynomial equation with the following solution:
\eqb
\kf=\frac{1}{2\alpha}\left(-1+\sqrt{1+\frac{12 \linj}{\lb \fex}} \right).
\label{ke1-fast-app}
\eqe
This is valid as long as $\alpha \kf< 1$ or equivalently if
\eqb
\fex > 6\frac{\linj}{\lb}.
\label{conf1-app}
\eqe
Equations (\ref{ke1-fast-app}) and (\ref{conf1-app}) should be compared to eqs.~(\ref{ke1-fast}) and (\ref{conf1}), respectively. 
Taking also into account the fast cooling condition, i.e. $\xi < 1$, we find 
\eqb
\fex > \frac{3}{4\lb \gamma_0 (1+4\linj \gamma_0)}
\label{conf2-app}
\eqe
Conditions (\ref{conf1-app}) and (\ref{conf2-app}) set a lower limit on $\fex$, which is written as
\eqb
\fex > \max\left(\frac{3}{4\lb \gamma_0 (1+4\linj \gamma_0)}, 6\frac{\linj}{\lb} \right).
\eqe
This is only slightly different than the one derived in Sect.~2.2.2 (see eq.~(\ref{conf4})).
\item for  $\alpha \kf > 1$, electrons may cool on the EC photons as well as on higher SSC photon generations. We are interested in the 
case where the $\ell_{\rm ssc, \nt-1} > \ell_{\rm EC}$. In this asymptotic limit, the multi-wavelength photon spectrum will be dominated by the highest
SSC photon generation. Thus, the features of the Compton catastrophe will not be hidden by a strong EC component. We may then write
\eqb
\ell_{\rm tot} \approx \ell_{\rm ssc, \nt-1} = \lb \left(\alpha \kf\right)^{\nt}.
\label{ltot_aprox3}
\eqe
In this limit, the energy density of EC scattered photons should not have an
obvious effect on the electron steady-state distribution. We find, indeed, that the solution to the steady-state eq.~(\ref{general-app})
is identical  to eq.~(\ref{ke2-fast}), namely
\eqb
\kf=\frac{1}{\alpha}\left(\frac{3\linj}{\lb}\right)^{1/(\nt+1)}.
\label{ke2-fast-app}
\eqe
The conditions $\ell_{\rm ssc, \nt-1} > \ell_{\rm EC}$ or, equivalently, $\lex < \lb(\alpha \kf)^{\nt-1}$ and $\xi < 1$ translate 
into the following two constraints on $\fex$ and $\linj$:
\eqb
\label{conf3-app}
\fex & <  & 1+\left(\frac{3\linj}{\lb}\right)^{\frac{\nt-1}{\nt+1}} \\
\linj &  >  & \frac{\lb}{3}\left(\frac{3}{4\lb \gamma_0}\right)^{\frac{\nt+1}{\nt}},
\label{conf4-app}
\eqe
where the upper limit on $\fex$ is slightly more stringent than before the inclusion of $u_{\rm EC}$.
\end{itemize}
Summarizing, we showed in detail that the inclusion of the energy density of EC scattered photons in the electron kinetic equation
has an effect on the steady-state solutions derived only in the fast cooling regime and on the respective constraints. 
This is not unexpected, since electron cooling becomes important, by definition, in the fast cooling regime.
Quantitatively speaking though, these effects are not important.

\section[]{The Compton $Y$ parameter}
\label{definitions}
In Sect.~\ref{AR}, we derived the conditions, in terms of the various
compactnesses, that result in
the dominance of the inverse Compton emission over the synchrotron one.  However, the strength
of the inverse Compton emission is traditionally expressed in terms of the Compton $Y$ parameter.

In the relativistic limit, the Compton $Y$ parameter is defined as  \citep{rybicki79}
\eqb
Y= \frac{4}{3}\langle \gamma_0^2 \rangle \max(\tau_{\rm e}, \tau_{\rm e}^2),
\eqe
where $\frac{4}{3}\langle \gamma_0^2 \rangle$ is the average energy gain per scattering and
$\tau_{\rm e}$ is the optical depth for Thomson scattering, which is given by
\eqb
\tau_{\rm e} = \sth \tilde{n}_{\rm el}\rb.
\eqe
In the above equation, $\tilde{n}_{\rm el}\equiv \gamma_0 \nel(\gamma_0)$ and
is the only quantity entering in the definition of $Y$ that  depends on $\linj$ and $\lb$. 
In the slow and fast cooling regimes, we find, respectively, that $\tilde{n}_{\rm el}=\ks$ and $\kf/\gamma_0$.
 Here, $\ks$ and $\kf$ are given by eqs.~(\ref{ke-slow}), (\ref{ke1-fast}) and (\ref{ke2-fast}), respectively.
The Thomson optical depth  can be, thus, written as
\eqb
\tau_{\rm e} = \left\{ \begin{array}{ll}
                      \frac{3 \linj}{\gamma_0} & {\rm slow \ cooling} \\ 
                      & \\
                      \frac{3\fex}{8\gamma_0^2}\left(-1+\sqrt{1+\frac{12\linj/\lb}{\fex^2}} \right) & {\rm fast \ cooling} \\ 
                      & \\
                      \frac{3}{4\gamma_0^2}\left(\frac{3\linj}{\lb} \right)^{1/(\nt+1)} & {\rm fast \ cooling \ (CC \ limit)}
                      \end{array}
                      \right.
                      \eqe

{ If $\tau_{\rm e} \gg 1$ (or, in practice, $\tau_{\rm e} \gtrsim 10$) our analysis is not valid,
since other physical processes that we do not take into account, such as photon-photon absorption, become
important. It is interesting though, that even for $\tau_{\rm e}\lesssim 1-3$, there are parameters
that drive the system to inverse Compton catastrophe. For simplicity, let us consider the slow cooling case.
In this case, higher order SSC photon generations become more luminous
for $\linj > 1/4\gamma_0$ (see Sect.~\ref{AR}), which translates into $\tau_{\rm e} > 3/4$. Thus, in the slow cooling regime and 
for $3/4< \tau_{\rm e} \lesssim 1-3$ the system may be driven to the Compton catastrophe regime, while  being still
optically thin to scatterings.}

For the parameter regime that does not lead
to the inverse Compton catastrophe and for $\tau_{\rm e} < 1$, the $Y$ parameter is written as
\eqb
Y = \left\{ \begin{array}{ll}
                     4\linj \gamma_0 & {\rm slow \ cooling} \\ 
                      & \\
                      \frac{\fex}{2}\left(-1+\sqrt{1+\frac{12\linj/\lb}{\fex^2}} \right) & {\rm fast \ cooling} \\ 
                     \end{array}
                      \right.
                      \eqe
Note that both expressions can be derived by $Y = \ell_{\rm ic}/\lsyn \approx \ell_{\rm ssc, 1}/\lsyn$  (see eqs.~(\ref{lssc-slow}), (\ref{lsyn-slow}),
(\ref{lssc1-fast}) and (\ref{lsyn1-fast})); the approximation breaks down in the Compton catastrophe regime. In the absence of external radiation fiels ($\fex=1$), 
the $Y$ parameter in the fast cooling regime reduces to

\eqb
Y \approx  \left\{\begin{array}{ll}
\sqrt{\frac{3\linj}{\lb}}, & \frac{12\linj}{\lb} \gg 1 \\
& \\
\frac{3\linj}{\lb}, & \frac{12\linj}{\lb} \ll 1\\
 \end{array}
                      \right.
\eqe
By identifying $12\linj$ as $\epsilon_{\rm e}$ and $\lb$ as $\epsilon_{\rm B}$, 
the above expressions are equivalent to these commonly used in GRB literature (e.g. \citealt{sarietal96, sariesin01}) 
when the scatterings take place
in the Thomson regime (for the expression of $Y$ in the presence of Klein-Nishina
effects, see \citealt{nakaretal09}).
\section[]{Observed EC flux for mono-energetic electrons}
\label{app2}
In the slow cooling regime, the electron distribution is defined by the balance between
the injection and escape terms. As long as the escape term is independent from
energy, the stationary electron distribution has the same 
energy dependence as the injection term. 
In our analysis we focus on mono-energetic injection. 

Assuming that the electron distribution in the comoving frame
is also isotropic we may write $\nel(\gamma, \Omega') =\ks / (4 \pi) \delta(\gamma-\gamma_0)$,
where primed quantities are measured in the blob's frame, while quantities with the subscript ``obs'' are measured
in the observer's frame. 
By inserting the above expression into eq.~(5) in D95 we find the EC emissivity in the comoving frame:
\eqb
j_{\rm EC}\left(\epsilon', \Omega' \right) = \frac{c  \sth u_{\rm ex}  \ks \gamma_0}{4\pi}\left(\frac{\epsilon'}{\eex} \right)^2\!\!
\left(\frac{\Gamma\eex(1+\mu')}{\epsilon'} \right)^{3/2}\!\!\delta\left(\epsilon' -\epsilon'_0\right),
\eqe
where  we used the property $\delta(g(x))= \sum_i \delta(x-x_i)/|g'(x_i)|$ with $x_i$ being the roots of  $g(x)=0$, and
\eqb
 \epsilon'_0 \equiv \gamma_0^2 \Gamma \eex (1+\mu'). 
\eqe
Substitution of $j_{\rm EC}\left(\epsilon', \Omega' \right)$ into eq.~(1) in D95 results in
\eqb
F_{\rm EC}^{\rm obs}(\eobs, \Omega_{\rm obs})  = F_0 
\left(\frac{\Dop  (1+\mobs)}{1+\beta}\right)^{3/2} \!\!\left(\frac{\eobs(1+z)}{\Dop \eex} \right)^{1/2}\!\!
\delta \left(\eobs - \epsilon_{\rm EC} \right).
\eqe
where we made use of the relation
\eqb
\Gamma \left(1+ \frac{\mobs-\beta}{1-\beta\mobs}\right)= \Dop \frac{1+\mobs}{1+\beta},
\eqe
 and
 \eqb
 F_0 & = & \frac{\Dop^4 \vb \sth c}{4\pi D_{\rm L}^2} u_{\rm ex} \ks \gamma_0 \\
 \epsilon_{\rm EC} & \equiv & \frac{\Dop^2 \eex \gamma_0^2 (1+\mobs)}{(1+z)(1+\beta)}.
 \eqe
\section[]{Expressions for the loss and injection operators used in the numerical calculations}
\label{app-num}
The numerical results presented in Sect.~\ref{NR} were obtained using a
numerical code that solves the coupled partial differential equations that govern
the evolution of photons and {relativistic} electrons\footnote{We do not distinguish between electrons and positrons.
We treat them as one particle population and we commonly refer to both as electrons.} in both energy and time \citep{mastkirk95}.
It is more convenient to rewrite eqs. (\ref{ne1}) and (\ref{ng1}) in dimensionless form, where
time is in units of the crossing time ($\tau=t c/\rb$), the photon and electron energy is in units of $\mel c^2$
and the particle number densities are normalized as
\eqb
\hat{n}_{\rm el}(\gamma, \tau) & = &  \nel (\gamma, \tau) \sth \rb \\
\hat{n}_{\gamma}(x, \tau) & = &  \nph(x, \tau) \sth \rb,
\eqe
where $\sth$ is the Thomson cross section.
In what follows, we present the expressions of the loss and source operators that appear in eqs.~(\ref{ne1})-(\ref{ng1}), after
transforming them into a dimensionless form. \\ \\
{\textbullet \sl Synchrotron radiation}\\
\eqb
\hat{Q}_{\gamma}^{\rm syn}(x,\tau) & = & \int_1^{\infty} d\gamma \  \nelhat(\gamma, \tau) \hat{j}_{\rm syn}(x,\gamma), \\
\hat{\mathcal{L}}_{\rm e}^{\rm syn}(\gamma, \tau) &  = & \frac{\partial}{\partial \gamma}\left(\nelhat(\gamma, \tau) \int_{\rm 0}^\infty dx \hat{j}_{\rm syn}(x,\gamma)\right),
\eqe
where the single particle synchrotron emissivity is 
\eqb
\label{syn-emissivity}
\hat{j}_{\rm syn}(x,\gamma) &  = & \frac{\sqrt{3} q_{\rm e}^3 B \sin \theta}{h x \mel c^2}\frac{\rb}{c} F\left(\frac{x}{x_{\rm c}}\right) \\
x_{\rm c} & = & \frac{3}{2}\sin \theta b \gamma^2,
\eqe
where $F(z)$ is defined in eq.~(6.31c) of \cite{rybicki79}, $\theta$ is the pitch angle between the magnetic field line
and the particle's velocity, and $b=B/\Bcr$.\\
{\textbullet \sl Synchrotron self-absorption} \\ \\
We treat synchrotron self-absorption only as a sink term for photons, while we neglect the respective energy gain term in the electron
equation. The loss term in the photon equation is written as 
\eqb
\hat{\mathcal{L}}_{\gamma}^{\rm ssa}(x, \tau) = R {\alpha_{\rm ssa}} \nphhat(x, \tau).
\eqe
In the above equation ${\alpha_{\rm ssa}}$ is the synchrotron self-absorption coefficient (c.f. eq.~(6.50) from \cite{rybicki79})
\eqb
{\alpha_{\rm ssa}} = \frac{c^3 h^3}{8 \pi (\mel c^2)^3}\frac{1}{\sth R^2 x^2} \int_1^{\infty} d\gamma \gamma^2 \frac{\partial}{\partial \gamma}\left(\frac{\nelhat(\gamma, \tau)}{\gamma^2} \right) 
\hat{j}_{\rm syn}(x, \gamma)
\eqe
 where $\hat{j}_{\rm syn}$ is defined in eq.~(\ref{syn-emissivity}). \\
{\textbullet \sl Inverse Compton scattering}\\
The energy injection term for photons is given by
\eqb
\hat{\mathcal{Q}}_{\gamma}^{\rm ics}(x_1,\tau) & = & \int d\gamma \nelhat(\gamma, \tau) \int dx\hat{j}_{\rm ics}(x, x_1, \tau),
\label{ics}
\eqe
where 
\eqb
\hat{j}_{\rm ics}(x, x_1,\tau)  = \frac{3}{4}\frac{x_1}{x}\frac{\nphhat(x)}{\gamma^2}F_{\rm C}(q, G).
\eqe
In the above, function $F_{\rm C}$ is given by eq.~(2.48) in \cite{blumenthalgould70}:
\eqb
F_{\rm C}(q, G) = 2 q \ln q + (1+2q)(1-q) + \frac{1}{2}\frac{\left(G q\right)^2}{(1+G q)}(1-q),
\eqe
where 
\eqb
G  & = & 4 \gamma x \\
q & = & \frac{E_1}{G(1-E_1)} \\
E_1 & = & \frac{x_1}{\gamma}.
\eqe
The electron energy loss term is split into two parts, 
$\hat{\mathcal{L}}_{\rm e}^{\rm ics}=\hat{\mathcal{L}}_{\rm e}^{\rm ics, T} + \hat{\mathcal{L}}_{\rm e}^{\rm ics, KN}$. 
For collisions taking place in the Thomson regime
the loss term is given by
\eqb
\hat{\mathcal{L}}_{\rm e}^{\rm ics, T}(\gamma, \tau) & = & \frac{4}{3}\frac{\partial}{\partial \gamma}\left(\gamma^2 \nelhat(\gamma, \tau)U_{\rm ph, T}\right) \\
U_{\rm ph, T}(\tau) & = & \int_0^{3/4\gamma} dx x \nphhat(x, \tau)
\eqe
while for scatterings occuring in the Klein-Nishina regime we may write

\eqb
\hat{\mathcal{L}}_{\rm e}^{\rm ics, KN}(\gamma, \tau) \approx \frac{3}{8}\frac{\nelhat(\gamma,\tau)}{\gamma}\int_{3/4\gamma}^{\infty} dx\frac{\nphhat(x,\tau)}{x}\left(\ln(4x\gamma) -\frac{11}{6}\right)
\eqe
where we made use of eq.~(2.57) in \cite{blumenthalgould70}. The above expression
for the inverse Compton losses  is based on the assumption that 
the electron losses all its energy in one collision while in the Klein-Nishina regime. \\ \\
{\textbullet \sl Photon-photon pair production} \\
The sink term for photons is written as
\eqb
\hat{\mathcal{L}_{\gamma}^{\gamma \gamma}}(x, \tau) = \nphhat(x, \tau) \int_{0}^{\infty} dx' \ \nphhat(x', \tau) R_{\gamma \gamma}(x x'),
\eqe
where for the reaction rate $R_{\gamma \gamma}$ we use the approximate expression (4.7) of \cite{coppiblandford90}
\eqb
R_{\gamma \gamma}(w) \approx 0.652 \frac{w^2-1}{w^3}\ln w \ H\left(w-1\right),
\label{Rgg}
\eqe
where $w =xx'$ and $H(z)$ is the usual Heaviside function. Assuming that the 
energy of the absorbed photon is shared between the electon and positron, the respective source
term for pairs becomes
\eqb
\hat{\mathcal{Q}_{\rm e}^{\gamma \gamma}}(\gamma, \tau) = 4 \nphhat(2\gamma, \tau)  \int_{0}^{\infty} dx' \ \nphhat(x', \tau) R_{\gamma \gamma}(2\gamma x'),
\eqe

 \section[]{Repeated scatterings of a fixed black-body photon field}
\label{app-comp}
We showed that, in the pure SSC case, the photon spectrum in the inverse Compton catastrophe limit
can be described by a broken power-law, with the spectral index below the 
break ($\beta_{\rm cc}$) depending on $\linj$, among
other parameters. We found that typically $\beta_{\rm cc} \gtrsim 0.6$, while only  for unphysically
high $\linj$, the photon spectra become harder. This is in contrast with one of the findings
in \cite{zdziarskilamb86} -- hereafter, ZL86, where photon spectra with $\beta \sim 0.3$ were obtained for $\linj$ values
lower than these adopted in this study. 

To test if there is a physical reason
for this discrepancy we 
calculate the photon spectra using
the same parameter values  as in Fig.~1 in ZL86: $B=0$, $R=8\times 10^6$~cm, $\gamma_{\min}=1$, $\gamma_{\max}=600$, $s=2.2$.
Our definition for the electron compactness has an extra factor $4\pi$ in the denominator (see eq.~(\ref{leinj})). Thus,
$\linj=\ell/(4\pi)=2.4$, for $\ell=30$. Moreover, the seed photon field  is assumed to be a black-body
with $T_{\rm BB} \simeq x_{\rm soft}/2.7 k \simeq 4\times 10^6$~K and compactness 
$\ell_{\rm soft}= \ell (L_{\rm soft}/L) = 0.35$, where $L_{\rm soft}/L=1/85$.

\begin{figure}
\centering
\resizebox{\hsize}{!}{\includegraphics{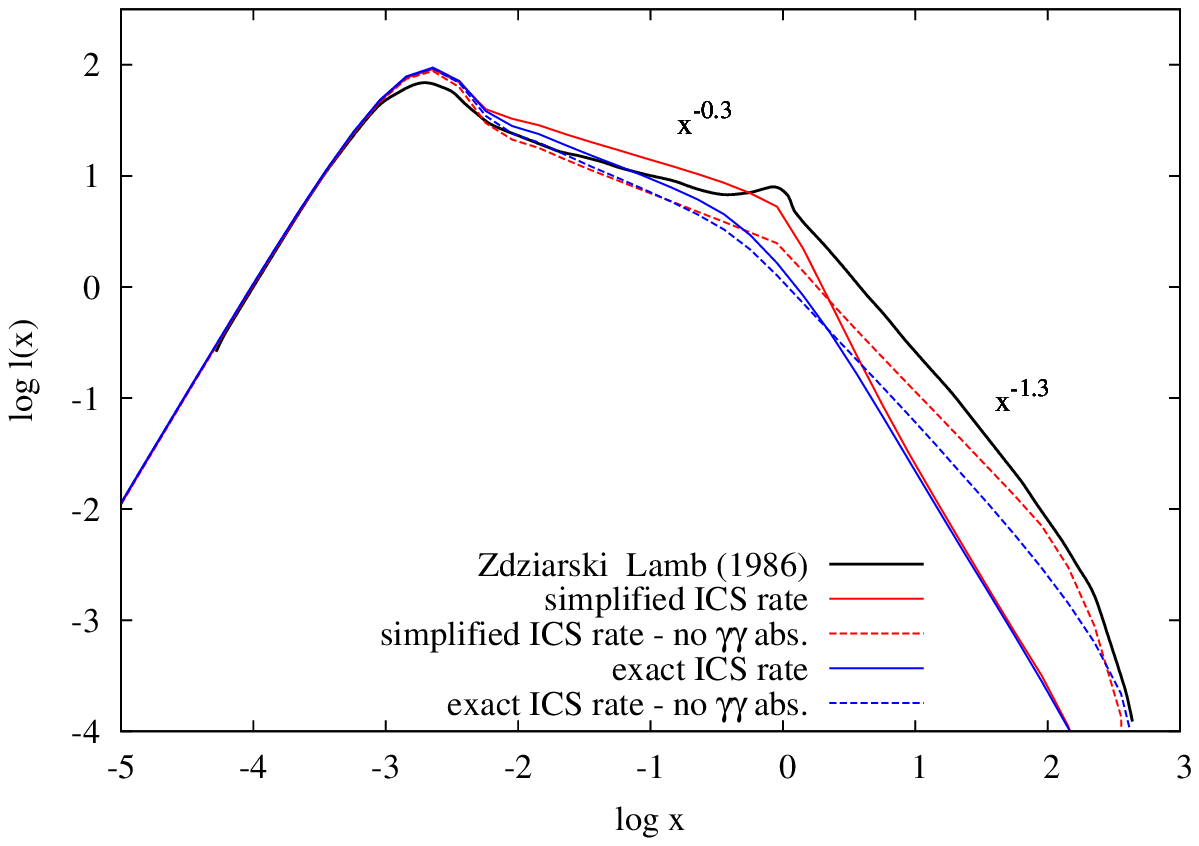}}
\caption{Photon spectra obtained for the same parameters used in Fig.~1 by ZL86. The spectrum
of ZL86 is shown with a black line, while our results are shown as red and blue curves.
Different lines have following meaning: simplified expression for the IC scattering
rate (red lines), exact expression for the IC rate from \citealt{blumenthalgould70} (blue lines), 
$\gamma \gamma$ absorption included (solid lines), $\gamma \gamma$ absorption neglected (dashed lines). }
\label{comparison}
\end{figure}

The comparison between the spectra is shown in Fig.~\ref{comparison}. The result of ZL86 is plotted with a black line, while
our numerical results are shown with red and blue lines. 
The red curves are obtained using a simplified expression for the Compton scattering rate (eq.~(44) in \citealt{mastkirk95}), which
is more close to the one used by ZL86. We also neglected scatterings
taking place in the Klein-Nishina regime. The blue curves are obtained 
using the full expression {(in the relativistic limit)} for the Compton scattering rate (see eq.~(\ref{ics})). 
Finally, to demonstrate the effect of photon-photon ($\gamma\gamma$) absorption on the spectra,
we artificially deleted the respective terms from the electron-photon equations (dashed lines).
A few remarks follow:
\begin{itemize}
\item for $x<1$, all spectra have similar spectral index ($\sim 0.3$).
\item the discontinuity found at $x\sim 1$  by ZL86 is a numerical artifact, that 
does not appear in our results.
\item for $x>1$, our spectra (with $\gamma\gamma$ absorption included) are steeper than the one
in ZL86. We note that ZL86 used a more crude expression for the $\gamma \gamma$ cross section than ours (see eq.~(\ref{Rgg})).
\item a comparison of the dashed lines and the black solid line shows
that for $x>1$ the red curves are more close to the result by ZL86, since they are obtained
using a similar expression for the Compton scattering rate.
\end{itemize}
Concluding, there are some differences at the high-energy part of the spectrum ($x>1$), which
can be understood in terms of differences in the $\gamma \gamma$ cross section and in the Compton
scattering rate. Yet, there is good agreement at the low-energy part of the spectrum ($x<1$). This
is harder than the spectrum we typically obtain in the Compton catastrophe regime, whenever
the  synchrotron photons serve as the seed photons for scattering.

\bibliographystyle{mn2e} 
\bibliography{ECgrb}
\end{document}